\newif\ifsubmission
\submissionfalse

\ifsubmission
\documentclass[acmsmall, screen, review]{acmart}
\else
\documentclass[letterpaper,11pt]{article}
\usepackage[margin=1in]{geometry}
\usepackage{hyperref}
\hypersetup{colorlinks,citecolor=blue}
\fi

\usepackage[ruled,linesnumbered,noend]{algorithm2e}
\ifsubmission\else\usepackage{amsmath}\fi
\usepackage{cleveref}[capitalize]
\newtheorem{theorem}{Theorem}

\usepackage{enumitem}
\usepackage{listings}
\usepackage{caption}
\usepackage{subcaption}
\usepackage{multirow}
\usepackage{tikz}
\usepackage{ulem}
\usetikzlibrary{arrows,automata,positioning}
\setitemize{leftmargin=*,topsep=0pt}
\setenumerate{leftmargin=*}
\definecolor{codegreen}{rgb}{0,0.6,0}
\definecolor{codegray}{rgb}{0.5,0.5,0.5}
\definecolor{codepurple}{rgb}{0.58,0,0.82}
\definecolor{backcolour}{rgb}{0.97,0.97,0.97}

\usepackage{filecontents}
\usepackage{url}

\newcommand{\tool}{\ensuremath{\texttt{Cr\^epe}}\xspace}

\newcommand{\ignore}[1]{}

\newcommand{\theories}{\mathcal{T}}
\newcommand{\theory}{T}
\newcommand{\proofrule}{\tau}
\newcommand{\proofstep}{\psi}
\newcommand{\formula}{\varphi}

\newcommand{\stad}[1]{\##1}
\newcommand{\exad}[1]{\&#1}
\newcommand{\chrad}[1]{\$#1}
\newcommand{\forad}[1]{\%#1}

\newcommand{\varid}{\textsf{Imm}\xspace}
\newcommand{\nodeid}{\textsf{NodeID}\xspace}
\newcommand{\arglist}{\textsf{Pointers}\xspace}
\newcommand{\eq}{\textsf{Eq}\xspace}
\newcommand{\stepid}{\textsf{StepID}\xspace}
\newcommand{\categoryid}{\textsf{CatID}\xspace}
\newcommand{\ruleid}{\textsf{RuleID}\xspace}
\newcommand{\premise}{\textsf{Premises}\xspace}

\newcommand{\result}{\textsf{Result}}
\newcommand{\premisesShort}{\textsf{Prems}}
\newcommand{\resultShort}{\textsf{Res}}
\newcommand{\validsteps}{\textsf{D}}

\newcommand{\mset}[1]{\texttt{multiset}(#1)}

\newcommand{\derivativeFunction}{\delta}
\newcommand{\derivative}[2]{\derivativeFunction(#1,#2)}
\newcommand{\hasblankFunction}{E}
\newcommand{\hasblank}[1]{\hasblankFunction(#1)}
\newcommand{\normalizeFunction}{\texttt{normalize}}
\newcommand{\normalize}[1]{\normalizeFunction(#1)}
\newcommand{\Sync}[3]{\text{Sync}(#1,#2,#3)}
\newcommand{\alphabet}{\Sigma}
\newcommand{\head}[1]{\texttt{head}(#1)}
\newcommand{\foot}[1]{\texttt{foot}(#1)}
\newcommand{\body}[1]{\texttt{body}(#1)}
\newcommand{\reduceFunction}{\texttt{reduce}}
\newcommand{\reduce}[1]{\reduceFunction(#1)}
\newcommand{\updateFunction}{\texttt{update}}
\newcommand{\update}[1]{\updateFunction(#1)}

\newcommand{\regexEmpty}{\emptyset}
\newcommand{\regexBlank}{\epsilon}
\newcommand{\regexUnion}{|}
\newcommand{\totalOrder}{R}
\newcommand{\identical}{==}
\newcommand{\similar}{\equiv}

\newcommand{\treesize}[1]{\texttt{size}(#1)}

\newcommand{\alphabetsize}{n}
\newcommand{\maxSyncLength}{\nu}
\newcommand{\syncTableLength}{\xi}
\newcommand{\termTableLength}{\chi}
\newcommand{\stepTableLength}{\pi}
\newcommand{\termTable}{M_t}
\newcommand{\syncTable}{M_s}
\newcommand{\formulaTable}{M_f}
\newcommand{\stepTable}{M_p}

\newcommand{\charNode}{\textsf{Char}\xspace}
\newcommand{\emptyNode}{\textsf{Empty}\xspace}
\newcommand{\blankNode}{\textsf{Blank}\xspace}
\newcommand{\unionNode}{\textsf{Union}\xspace}
\newcommand{\concatNode}{\textsf{Concat}\xspace}
\newcommand{\starNode}{\textsf{Star}\xspace}
\newcommand{\epsilonNode}{\textsf{Epsilon}\xspace}
\newcommand{\derivativeNode}{\textsf{Derivative}\xspace}

\newcommand{\syncNode}{\textsf{Sync}\xspace}

\newcommand{\equivAlg}{\texttt{EQUIV}}
\newcommand{\proofAlg}{\texttt{PROOF}}

\newcommand{\researchQ}[1]{\textbf{Q#1}}

\begin{document}

\date{}

\title{Coinductive Proofs of Regular Expression Equivalence in Zero Knowledge}

\ifsubmission
\else
\author{
John Kolesar\\
Yale University\\
{\footnotesize\url{john.kolesar@yale.edu}}
\and
Shan Ali\\
Yale University\\
{\footnotesize\url{shan.ali@yale.edu}}
\and
Timos Antonopoulos\\
Yale University\\
{\footnotesize\url{timos.antonopoulos@yale.edu}}
\and
Ruzica Piskac\\
Yale University\\
{\footnotesize\url{ruzica.piskac@yale.edu}}
}
\fi

\maketitle

\ifsubmission
\else
\begin{abstract}

Zero-knowledge (ZK) protocols enable software developers to provide proofs of their programs' correctness to other parties without revealing the programs themselves.
Regular expressions are pervasive in real-world software,
and zero-knowledge protocols have been developed in the past for the problem of checking whether an individual string appears in the language of a regular expression, but no existing protocol addresses the more complex PSPACE-complete problem of proving that two regular expressions are equivalent.

We introduce \tool, the first ZK protocol for encoding regular expression equivalence proofs and also the first ZK protocol to target a PSPACE-complete problem.
\tool uses a custom calculus of proof rules based on regular expression derivatives and coinduction, and we introduce a sound and complete algorithm for generating proofs in our format.
We test \tool on a suite of hundreds of regular expression equivalence proofs.
\tool can validate large proofs in only a few seconds each.

\end{abstract}
\fi


\section{Introduction}
\label{sec:intro}

Software verification is the process of translating a program into a mathematical formalism and then constructing a proof that the formalized program satisfies some specification.
Traditional verification tools do not allow programmers to hide the implementation details of their software from other parties.
The tools require all relevant information to be fully visible to all parties involved in the verification process, so they are inadequate for
verification problems involving third-party libraries, cloud-based services, or other proprietary software:  the owners of the software cannot share the evidence of the program's correctness with other parties without losing their privacy.

To make verification of private programs possible, recent research at the intersection of cryptography and formal verification has focused on the development of zero-knowledge (ZK) protocols for encoding proofs of programs' correctness~\cite{luo2022proving,luick2023zksmt,cryptoeprint:2024/267}.
A zero-knowledge protocol is a cryptographic method of communication that allows one party to demonstrate knowledge of a secret to another party without revealing the secret~\cite{goldwasser2019knowledge,goldreich1991proofs}.
The ZK program verification process involves two parties:  the \textit{prover} who owns the private program and the \textit{verifier} whom the prover wants to convince of the program's correctness.
(In the domain of ZK protocols, the terms ``prover'' and ``verifier'' have different meanings than they do in the domain of non-cryptographic formal verification.)
The prover constructs a proof of the program's correctness offline and then engages in an interactive zero-knowledge protocol with the verifier.
During the protocol, the prover provides evidence that suffices to convince the verifier that the program is correct but does not reveal additional information about the program's implementation.

To be efficient and scalable, ZK protocols must be tailored for specific domains.
Existing ZK verification protocols have targeted Boolean logic~\cite{luo2022proving}, quantifier-free first-order logic with EUF and LIA~\cite{luick2023zksmt}, and Lean's calculus of dependent types~\cite{cryptoeprint:2024/267},
but currently no ZK protocol exists for the problem of regular expression equivalence.
Regular expressions are a common formalism for defining string patterns,
and their uses in modern software are pervasive~\cite{brzozowski1962survey,sommer2003enhancing,van2006high}.
In particular, the applications of regular expression equivalence
include correctness proofs for compiler optimizations~\cite{Kozen2000CertificationOC}, translation validation~\cite{foster2015coalgebraic}, and query optimization~\cite{calvanese1999rewriting}.
Also, individual regular expressions for tasks such as intrusion detection and prevention for network packets~\cite{zhao2020achieving} can be sensitive intellectual property, so programmers can benefit from a ZK protocol for reasoning about the semantics of private regular expressions.

Importantly, we cannot reformulate the problem of regular expression equivalence in terms of problems that existing ZK protocols handle.
Existing ZK regular expression protocols~\cite{zhang2024zombie,luo2023privacy,raymond2023efficient,angel2023reef} reason only about
the problem of matching individual strings against regular expressions.
Regular expression equivalence is a PSPACE-complete problem~\cite{stearns1985equivalence,stockmeyer1973word},
and string matching is at most NP-complete if generalized~\cite{alfred2014algorithms}, so we cannot reduce regular expression equivalence to string matching unless P=PSPACE or NP=PSPACE.
In fact, no existing practical ZK protocol targets a PSPACE-complete problem at all,
even though
it has been known for decades that the complexity class IP (the set of all problems that can be solved with interactive ZK proofs) is equal to PSPACE~\cite{shamir1992ip}.

In this paper,
we introduce the first practical ZK protocol
for the problem of regular expression equivalence.
More specifically, we developed a tool called
\tool (Cryptographic Regex Equivalence Proof Engine) that
validates regular expression equivalence proofs in ZK.
When two parties communicate using \tool,
the owner of a hidden proof that two regular expressions are equivalent can convince the other party that the proof is valid without leaking the regular expressions or the proof.
Also, we introduce a new decision procedure to generate the proofs that \tool validates.
We cannot rely on an existing solver to generate proofs for us:
older tools have only limited support for regular expression proofs.
Modern SMT solvers can reason about other formalisms effectively,
but there is no well-established format for providing proof certificates about regular expression equivalence analogous to the existing formats for Boolean logic, EUF, and LIA~\cite{christ2012smtinterpol}.
Given two equivalent regular expressions as input, our decision procedure constructs an equivalence proof for them in a custom calculus of proof rules that we introduce.
Instead of converting regular expressions into state machines or other objects, as standard solvers do~\cite{zheng2017z3str2,berzish2021smt},
our rules apply coinduction to reason about regular expressions directly, eliminating the need to validate a conversion between different formats.
We prove that our rules are sound and complete for regular expression equivalence.

In summary, we make four main research contributions:

\begin{enumerate}
    \item\textbf{ZK Protocol.} We introduce \tool, the first zero-knowledge protocol designed to validate regular expression equivalence proofs and also the first practical ZK protocol to target a PSPACE-complete problem.
    \item\textbf{Proof Generator.} We develop a backend for \tool that generates regular expression equivalence proofs in a custom calculus, and we prove the calculus sound and complete, which is of independent interest.
    \item\textbf{Proof Generator Evaluation.} We test our custom proof generator on a suite of commonly-used regular expression benchmarks from FlashRegex~\cite{li2020flashregex}.
    The regular expressions range from 2 to 60 characters in length.
    Our proof generator scales to handle large, complex inputs, and it generates equivalence proofs successfully for almost all of the equivalent pairs in the suite.
    \item\textbf{ZK Protocol Evaluation.} We test \tool on a suite of hundreds of equivalence proofs from our custom proof generator.
    \tool can validate 84.39 percent of the proofs in 15 seconds or less.
    Also, we run \tool on the same suite with alternative configurations that change the amount of information that it leaks.
    Our default settings provide strong safety guarantees and incur only a 35\% median slowdown relative to a version that leaks more information.
\end{enumerate}

\section{Motivating Example}
\label{sec:redos}

Unsafe usage of regular expressions opens a window for attacks that flood a network server with unwanted slow traffic, making it inaccessible.
A \textit{denial-of-service attack} against a regular expression (ReDoS or simply DoS) is the use of a malicious input to cause a regular expression matching algorithm to run in super-linear time~\cite{bhuiyan2024sok}.
ReDoS attacks are possible because the standard approach for regular expression matching is to convert the regular expression into a nondeterministic finite automaton (NFA) and to evaluate the NFA on the input string~\cite{davis2018impact}.
NFA evaluation permits unlimited branching, and the number of branching paths can become polynomial or exponential in the size of the input string.
This branching can cause serious harm in practice:
in 2019, Cloudflare suffered a 27-minute global outage because the NFA-based string matching in their code ran in exponential time on a specific regular expression~\cite{cloudflare2019,cloudflare2020}.
Even regular expressions for mundane tasks such as validating e-mail addresses can be vulnerable to exponential branching:

\ifsubmission
\else
\begin{center}
\texttt{\^{}([0-9a-zA-Z]([-.\textbackslash w]*[0-9a-zA-Z])*@\\([0-9a-zA-Z][-\textbackslash w]*[0-9a-zA-Z]\textbackslash .)+[a-zA-Z]\{2,9\})\$}
\end{center}
\fi
This regular expression for validating e-mail addresses comes from RegExLib, a large public online database of regular expressions~\cite{RegExLib541}.
The sub-expression \texttt{([-.\textbackslash w]*[0-9a-zA-Z])*} can cause string matching to run in exponential time because it contains a \texttt{*} quantifier inside another \texttt{*}-quantified sub-expression.

To illustrate the vulnerability, we will use the simpler regular expression $(a^*a)^*$, which has the same general structure and vulnerability as the e-mail address example.
A representation of $(a^*a)^*$ as an NFA appears in Figure~\ref{fig:state-machine-two-stars}.
The NFA has multiple options for processing $aa$ when starting from $v_1$.
For one option, it takes the $\regexBlank$ transition to $v_2$, consumes one $a$ with $v_2$'s self-edge, consumes the second $a$ by moving to $v_3$, and then takes the $\regexBlank$ transition back to $v_1$.
Another option is to move from $v_1$ to $v_2$, then to $v_3$, and back to $v_1$ to consume one $a$, repeating the process to consume a second $a$.
(There are more possible paths, but we only need to consider these two.)
The NFA will branch and explore both paths whenever it is at $v_1$ and encounters two consecutive $a$ characters.

\ifsubmission
\else
\begin{figure}
\centering
    \begin{subfigure}[t]{0.45\textwidth}
        \centering
        \scalebox{1}{
        \begin{tikzpicture}[shorten >=1pt,node distance=2cm,on grid,auto]
            \tikzstyle{every state}=[fill={rgb:black,1;white,10}]
    
            \node[state,initial,accepting]   (v_1)                 {$v_1$};
            \node[state] (v_2) [right of=v_1]  {$v_2$};
            \node[state] (v_3) [below of=v_2]  {$v_3$};
    
            \path[->]
                (v_1) edge              node {$\regexBlank$}  (v_2)
                (v_2) edge  [loop right]  node {a}  (v_2)
                (v_2) edge              node {a}  (v_3)
                (v_3) edge    node {$\regexBlank$} (v_1);
        \end{tikzpicture}
        }
        \caption{A state machine for $(a^*a)^*$}
        \label{fig:state-machine-two-stars}
    \end{subfigure}
\hfill
    \begin{subfigure}[t]{0.45\textwidth}
        \centering
        \scalebox{1}{
        \begin{tikzpicture}[shorten >=1pt,node distance=2cm,on grid,auto]
              \tikzstyle{every state}=[fill={rgb:black,1;white,10}]
            
              \node[state,initial,accepting]   (v_0)                 {$v_4$};
            
              \path[->]
              (v_0) edge  [loop right]  node {a}  (v_0);
        \end{tikzpicture}
        }
        \caption{A state machine for $a^*$}
        \label{fig:state-machine-one-star}
    \end{subfigure}
\caption{State machines for motivating example}
\end{figure}
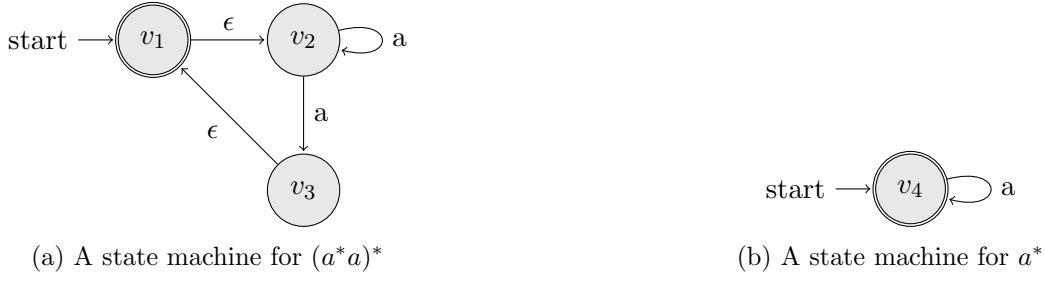
\fi

For an input string that consists of $2m$ copies of $a$ followed by $b$, where $m$ is a large positive integer, the NFA has at least $2^m$ paths to explore.
Each path will terminate with a non-accepting result when it reaches the $b$ at the end.
The NFA needs to examine all of the paths to confirm that $(a^*a)^*$ does not accept the input string, so it will run in exponential time.
A similar string works as a DoS attack for the original e-mail address regular expression.

There exists an equivalent regular expression that is immune to the attack, namely $a^*$.
An NFA
for $a^*$ appears in Figure~\ref{fig:state-machine-one-star}.
The machine has one state and only one path that it can take to consume a character, so branching is impossible.
If the machine ever reads a character other than $a$, the execution path terminates and rejects the input string.
Therefore, the NFA for $a^*$ will not run in super-linear time on any input.

By using \tool and our sound and complete custom calculus of rules, we can show in ZK that $(a^*a)^*$ and $a^*$ are equivalent.
The protocol involves two parties:  the \textit{prover} who owns both regular expressions and the \textit{verifier} whom the prover wants to convince that the regular expressions are equivalent without revealing the regular expressions.
First, the prover needs to create an equivalence proof offline.
The proof relies on derivatives~\cite{brzozowski1964derivatives} and coinduction~\cite{10.1145/1516507.1516510}.
Informally, derivatives represent the changes that happen in a regular expression's behavior as it reads characters from an input string.
For instance, the derivative of $ab\regexUnion ac$ with respect to $a$ is $b\regexUnion c$.
Coinduction is a proof technique that allows us to take advantage of cyclic patterns in the regular expressions' derivatives.
A proof of the equivalence of $(a^*a)^*$ and $a^*$ appears in Table~\ref{tab:intro-proof}.
In the first few steps, we examine the derivatives of $(a^*a)^*$ and $a^*$.
Later, we find that $(a^*a)^*$ and $a^*$ retain their behavior when they read an additional $a$ from an input string, so we apply coinduction to the cycle.
We know from our coinductive step that $(a^*a)^*$ and $a^*$ agree on any number of repetitions of $a$, and neither regular expression accepts strings with characters other than $a$, so the two regular expressions must be equivalent.
We explain derivatives and coinduction in more detail in Section~\ref{sec:preliminaries}.
A more detailed explanation of the proof appears in Section~\ref{sec:protocolDesign}.

After generating an equivalence proof, the prover can allow the verifier to validate the proof in ZK.
When the prover and verifier communicate using \tool,
their interaction consists of a series of steps.
In each step, the prover provides the verifier with evidence that an individual proof step is valid.
The evidence corresponds to a single row of the proof step table:
the verifier learns that the proof step is valid but cannot see the specific operations that the step performs.
When the prover and verifier finish validating every individual step and also confirm some other necessary properties such as the absence of cyclic pointers (Section~\ref{sec:protocolDesign}), the verifier knows that the proof as a whole is valid.
We discuss our ZK operations in more detail in Section~\ref{sec:zk implementation}.

\ifsubmission
\else
\begin{figure}
\centering
\begin{tabular}{|@{\hskip 2pt}c@{\hskip 2pt}|@{\hskip 2pt}c@{\hskip 2pt}|@{\hskip 2pt}c@{\hskip 2pt}|@{\hskip 2pt}l@{\hskip 2pt}|}
    \hline 
    {\bf\stepid} & {\bf\ruleid} & {\bf\premise} & {\bf\result} \\
    \hline
    $\stad{1}$ & {\sf Derive} & & $\derivative{a}{(a^*a)^*}$ \\
    & & & $=(a^*a\regexUnion\regexBlank)(a^*a)^*$ \\
    \hline
    $\stad{2}$ & {\sf Derive} & & $\derivative{a}{(a^*a\regexUnion\regexBlank)(a^*a)^*}$ \\
    & & & $=(a^*a\regexUnion\regexBlank)(a^*a)^*$ \\
    \hline
    $\stad{3}$ & {\sf Derive} & & $\derivative{a}{a^*}=a^*$ \\
    \hline
    $\stad{4}$ & {\sf Epsilon} & & $\hasblank{(a^*a\regexUnion\regexBlank)(a^*a)^*}=\regexBlank$ \\
    \hline
    $\stad{5}$ & {\sf Epsilon} & & $\hasblank{(a^*a)^*}=\regexBlank$ \\
    \hline
    $\stad{6}$ & {\sf Epsilon} & & $\hasblank{a^*}=\regexBlank$ \\
    \hline
    $\stad{7}$ & {\sf Eq} & $\{\stad{4},\stad{6}\}$ & $\hasblank{(a^*a\regexUnion\regexBlank)(a^*a)^*}=\hasblank{a^*}$ \\
    \hline
    $\stad{8}$ & {\sf Eq} & $\{\stad{5},\stad{6}\}$ & $\hasblank{(a^*a)^*}=\hasblank{a^*}$ \\
    \hline

    $\stad{9}$ & {\sf SyncCycle} & $\{\stad{2},\stad{3}\}$ & $\Sync{a}{(a^*a\regexUnion\regexBlank)(a^*a)^*}{a^*}$ \\
    \hline
    $\stad{10}$ & {\sf Coinduction} & $\{\stad{4},\stad{9}\}$ & $\Sync{\regexBlank}{(a^*a\regexUnion\regexBlank)(a^*a)^*}{a^*}$ \\
    \hline
    $\stad{11}$ & {\sf SyncEmpty} & $\{\stad{10}\}$ & $(a^*a\regexUnion\regexBlank)(a^*a)^*=a^*$ \\
    \hline
    $\stad{12}$ & {\sf Cong} & $\{\stad{1},\stad{11}\}$ & $\derivative{a}{(a^*a)^*}=a^*$ \\
    \hline
    $\stad{13}$ & {\sf Cong} & $\{\stad{3},\stad{12}\}$ & $\derivative{a}{(a^*a)^*}=\derivative{a}{a^*}$ \\
    \hline
    $\stad{14}$ & {\sf Match} & $\{\stad{8},\stad{13}\}$ & $(a^*a)^*=a^*$ \\
    \hline
    \end{tabular}
    \captionof{table}{A proof of the equivalence of $(a^*a)^*$ and $a^*$ in \tool's format.  Some steps are omitted or simplified.  We use $\stad{}$ to denote the addresses of proof steps.}
    \label{tab:intro-proof}
\end{figure}
\fi

\section{Preliminaries}
\label{sec:preliminaries}

\subsection{Grammar}

The three main categories of entities in our formalism for \tool are strings ($s$), terms ($p$), and formulas ($\formula$).
We give inductive definitions for all three:

\begin{align*}
s ::=\ & \regexBlank \mid cs_1 \\
p ::=\ & \regexEmpty \mid \regexBlank \mid c \mid p_1p_2 \mid p_1\regexUnion p_2 \mid {p_1}^* \mid \hasblank{p_1} \mid \derivative{c}{p_1} \\
\formula ::=\ & p_1=p_2 \mid \Sync {s_1}{p_1}{p_2} \mid p_1\neq p_2 \mid \bot
\end{align*}

A string is simply a list of characters from an alphabet $\alphabet$.
Starting from the empty string $\regexBlank$, we build longer strings inductively by adding characters to the front of shorter ones.
Sometimes we write $sc$ to denote the string $s$ with $c$ added to the back.
Note that we use $c$ to denote the individual character $c$, the one-character string whose only character is $c$, and also the regular expression that accepts only the string $c$.
Additionally, we use $\regexBlank$ to denote both the empty string and the regular expression that accepts only the empty string.

Terms represent regular expressions and functions over regular expressions.
Every term defines a \textit{language}:  the set of strings that it accepts.
$\regexEmpty$ is the regular expression whose language is empty:  it accepts no strings.
$\regexBlank$ accepts the empty string and nothing else.
The regular expression $c$ accepts the one-character string $c$ and nothing else, where $c$ is a character from $\alphabet$.
$p_1\regexUnion p_2$ is the union of terms $p_1$ and $p_2$:  it accepts all strings accepted by either $p_1$ or $p_2$.
$p_1p_2$ is the concatenation of $p_1$ and $p_2$:  it accepts all strings that consist of a part that $p_1$ accepts followed by a part that $p_2$ accepts.
Lastly, $p_1^*$ accepts zero or more repetitions of $p_1$.
$\hasblank{p_1}$ and $\derivative c{p_1}$ are functions over terms, where $c$ is a character.
We say that a term is a \textit{regular expression} if it does not contain any AST nodes of the form $\hasblank p$ or $\derivative cp$.

A \textit{formula} $\formula$ is an assertion about terms that has a truth value.
Every proof step has a formula as its conclusion.
We support four predicates for formulas.
Equality ($p=q$) means that two terms have the same language.
The Sync predicate ($\Sync {s_1}{p_1}{p_2}$) serves as an indicator of incremental progress toward a proof that two terms are equivalent.
Informally, $\Sync {s_1}{p_1}{p_2}$ means that, if $p_1$ and $p_2$ disagree on a string that starts with $s_1$, they also disagree on a strictly shorter string.
We will explain the meaning of Sync in more depth in Section~\ref{sec:coinduction-rules}.
Inequality ($p_1\neq p_2$) is simply the negation of equality.
Lastly, bottom ($\bot$) indicates that a proof has reached a contradiction.
All of our proofs conclude $\bot$ in the final step.

\paragraph{Epsilon}
As part of our formalism, we include an \textit{epsilon} function $\hasblankFunction$.
If $p$ is a term, then $\hasblank p$ indicates whether $p$ accepts the empty string $\regexBlank$.
More specifically, $\hasblank p$ returns $\regexBlank$ if $p$ accepts $\regexBlank$ and returns $\regexEmpty$ otherwise.
Having $\hasblankFunction$ return a term rather than a Boolean value allows us to avoid introducing an extra type into our formalism.

\paragraph{Derivatives}
\label{sec:derivatives}

If $p$ is a term and $c$ is a character, then $\derivative cp$ is the \textit{derivative} of $p$ with respect to $c$:
the term whose language is the set of strings $s$ such that $p$ accepts $cs$~\cite{brzozowski1964derivatives}.
We can generalize the definition of derivatives to strings once we have a definition for individual characters.
For the base case, $\derivative\regexBlank p=p$ for any $p$.
For the inductive case, $\derivative{sc}p=\derivative{c}{\derivative sp}$ for a string $s$ and character $c$.
Note that the derivative nodes for individual characters are the only derivative nodes that exist within our abstract syntax trees (ASTs):  we introduce this notation only for the sake of readability.

\subsection{Regular Expression Equivalence}
\label{sec:regex-equivalence}

Two regular expressions are equivalent if and only if they have the same language.
Regular expression equivalence is a decidable problem and is PSPACE-complete~\cite{stearns1985equivalence,stockmeyer1973word}.
Decision procedures for regular expression equivalence have existed for decades~\cite{kaplan1969regular,antimirov1995rewriting}, but the task of generating regular expression equivalence proofs that can be validated by parties other than the prover has received comparatively little attention.
From a certain perspective, this is unsurprising:  since the problem is PSPACE-complete, validating a proof of two regular expressions' equivalence is asymptotically just as costly as finding the proof in the first place.
However, when privacy concerns arise, the ability to validate a regular expression equivalence proof constructed by another party becomes valuable.

\paragraph{Custom Proof Format}
We use a custom backend to generate regular expression equivalence proofs for \tool.
We do not use an existing solver to generate proofs because existing SMT solvers provide only limited support for reasoning about regular expressions.
SMTInterpol~\cite{christ2012smtinterpol}, the solver used by the ZK protocol ZKSMT for proof generation~\cite{luick2023zksmt}, does not support regular expressions at all.
CVC5 can check whether an individual string appears in the language of a regular expression, but its rules for reasoning about regular expression equivalence are not complete~\cite{kuhne2021automatically}.
Existing solvers that are complete for regular expression equivalence are not suitable backends for \tool either.
The standard technique for comparing regular expressions to each other is to convert the regular expressions into state machines.
The SMT solver Z3 uses this approach~\cite{zheng2017z3str2,berzish2021smt}, and its algorithm is complete for regular expression equivalence~\cite{z3regex}.
However, Z3 cannot always provide full evidence for the correctness of the SAT/UNSAT answers that it returns.
Z3 does not support a full axiomatization of the theory of regular expressions in terms of rules for proof certificates~\cite{bjorner2025}.
When Z3 applies theory-specific reasoning for regular expression equivalence or other non-axiomatized theories, it represents the process in its proof certificates as a black-box term rewriting step~\cite{de2008proofs}.
All reasoning in a zero-knowledge proof needs to be explicit, so opaque term rewriting steps make Z3's proof certificates unusable in ZK.
Moreover, Z3 gained the ability to provide certificates for any regular expression pair whose equivalence it can prove only recently, in parallel to the development of \tool.
Before it received the extension, Z3's support for proof certificates about regular expression equivalence was even more limited.
Previously, when the user requested a proof rather than only a SAT/UNSAT answer, Z3 lost completeness for regular expression equivalence, and it managed to provide certificates only for simple inputs~\cite{bjorner2024}.

\tool's regular expression equivalence proofs do not involve state machine conversion.
Instead, they operate on regular expressions directly by using equations and algebraic data types.
If we used proofs based on state machines, the proof would need to contain evidence of the correctness of our conversion of each starting regular expression into a corresponding state machine.
Since our proofs deal with regular expressions directly, they eliminate the need to validate conversions between different formats.

\subsection{Coinduction}
\label{sec:coinduction-intro}

\textit{Coinduction} serves as the backbone of \tool's custom proof format.
Coinduction is a proof technique analogous to induction for reasoning about potentially infinite data structures.
Instead of using base cases and inductive steps to prove that a property holds for all finite structures of a specific type,
a coinductive proof constructs a \textit{bisimulation} between two objects to demonstrate that they uphold a property and continue to uphold the property after any number of reductions.
For our purposes, the two infinite objects being compared are the paths that two regular expressions' derivatives can take, and the property that they continue to hold after any number of reductions is the derivatives' equivalence.
We use the Sync predicate to construct the bisimulation gradually.

Coinduction is a more suitable technique for regular expression equivalence proofs than induction is.
Regular expressions are defined inductively, but the derivative function
for regular expressions does not make gradual progress toward a base case.
For a regular expression $p$ and character $c$, there is no guarantee that $\derivative cp$ will contain fewer AST nodes than $p$ itself.
For instance, $\derivative{a}{(ab)^*}$ is $b(ab)^*$, which is larger than $(ab)^*$.
Although the derivative function is not guaranteed to reach a terminating case, it is guaranteed to reach a fixed point or cycle eventually since the derivatives of a regular expression must fall into a finite number of equivalence classes~\cite{brzozowski1964derivatives}.
We can utilize the cycles for our coinductive proofs.
If the two regular expressions being compared behave equivalently up to the point where they reach a cycle, further repetitions of the cycle will not cause them to behave differently.
Consequently, when we find cycles in the derivatives of two regular expressions, we can work backward from the cycles to prune the search space until we know that the regular expressions align on all strings.

\subsection{Zero-Knowledge Proofs}

A zero-knowledge protocol is a method of communication between two parties, known as the \textit{prover} and the \textit{verifier}.
Both parties know a public predicate $P$, and the prover possesses a private \textit{witness} $w$.
The prover's goal is to demonstrate to the verifier that $w$ satisfies $P$ without revealing the value of $w$~\cite{goldwasser2019knowledge,goldreich1991proofs}.
The protocol may leak some information about $w$, but the verifier should not be able to make incremental progress toward recovering the entirety of $w$ by executing the protocol a large number of times.
In the context of \tool, the witness $w$ is a logical deduction showing that two regular expressions are equivalent.
The public predicate $P$ is the assertion that the deduction is valid.
Importantly, the word ``proof'' has two different senses in the domain of ZK verification.
Logical proofs and ZK proofs are separate concepts.
For our purposes, the prover wants to provide a zero-knowledge proof of the existence of a logical proof.

\tool is a \textit{commit-and-prove} ZK protocol.
Commit-and-prove protocols use a technique known as \textit{commitment} to conceal witnesses~\cite{canetti2002universally}.
Within a ZK proof, the verifier cannot see the underlying value of a committed object but can see that the results of all operations on the committed object are consistent with each other.
In our proofs for \tool, we perform addition, multiplication, and comparison operations on committed integers.
Our formalism is not tied to any specific commit-and-prove system,
but, for our implementation,
we use a recently developed VOLE-ZK backend~\cite{weng2021wolverine}.

\section{Protocol Design}
\label{sec:protocolDesign}

\paragraph{Proof Structure}

Many aspects of \tool's design resemble the designs of ZKSMT~\cite{luick2023zksmt} and zkPi~\cite{cryptoeprint:2024/267}, two other ZK protocols for encoding proofs.
An instance of \tool is a virtual machine with read-only memory, a set of checking instructions, and a list of proof steps.
Our proofs involve three types of inductively-defined structures:  terms, strings, and formulas.
We represent all three as ASTs, where each inductive constructor is a single AST node.
An instance of \tool includes three read-only tables for storing them.
The \textit{term table} $\termTable$ contains all of the terms in a proof.
Each entry in the table represents an AST node, and nodes store pointers to their immediate children within the table.
The \textit{string table} $\syncTable$ contains the strings used for Sync predicates.
We represent strings as singly-linked lists.
Each node stores a single character, namely the one at the front of the string, and also stores a pointer to the remainder of the list.
One entry in the string table is a null terminator that represents $\regexBlank$.
All chains of linked list nodes in $\syncTable$ ultimately lead to the null terminator.
The \textit{formula table} $\formulaTable$ contains the conclusions of all of the proof steps in the proof.
In our setup, formulas do not contain pointers to other formulas, but they can build on top of terms and strings.
In that event, a formula's entry in $\formulaTable$ contains pointers to entries from $\termTable$ and $\syncTable$.

\begin{table}
\centering
\begin{tabular}{|@{\hskip 3pt}c@{\hskip 3pt}|@{\hskip 3pt}c@{\hskip 3pt}|@{\hskip 3pt}c@{\hskip 3pt}|@{\hskip 3pt}l@{\hskip 3pt}|@{\hskip 3pt}c@{\hskip 3pt}|}
    \hline 
    {\sf\bf Addr.} & {\bf\nodeid} & {\bf\varid} & {\bf\arglist}  & {\bf\sf Meaning}\\
    \hline
    $\exad{0}$ & \emptyNode &   & $\{ \}$ & $\regexEmpty$ \\
    \hline
    $\exad{1}$ & \blankNode &  & $\{ \}$ & $\regexBlank$ \\
    \hline 
   $\exad{2}$ & \charNode &  $a$  &  $\{\}$ & $a$\\
   \hline 
   $\exad{3}$ & \starNode & &  $\{\exad{2}\}$ & $a^*$\\
   \hline 
   $\exad{4}$ & \concatNode &  &  $\{\exad{3},\exad{2}\}$ & $a^*a$\\
   \hline
   $\exad{5}$ & \starNode & &  $\{ \exad{4} \}$ & $(a^*a)^*$\\
   \hline
   $\exad{6}$ & \epsilonNode & &  $\{ \exad{3} \}$ & $\hasblank{a^*}$\\
   \hline
   $\exad{9}$ & \epsilonNode & &  $\{\exad{5}\}$ & $\hasblank{(a^*a)^*}$\\
   \hline
   $\exad{10}$ & \derivativeNode & $a$ &  $\{\exad{3}\}$ & $\derivative{a}{a^*}$\\
   \hline
   $\exad{11}$ & \unionNode & &  $\{\exad{4},\exad{1}\}$ & $a^*a\regexUnion\regexBlank$\\
   \hline
   $\exad{12}$ & \concatNode & &  $\{\exad{11},\exad{5}\}$ & $(a^*a\regexUnion\regexBlank)(a^*a)^*$\\
   \hline
   \end{tabular}
    \captionof{table}{Part of the term table $\termTable$ for the proof of the equivalence of $(a^*a)^*$ and $a^*$. We use $\exad{}$ to denote the addresses of terms.}
    \label{tab:expr-table-from-example}
\ifsubmission\else\vspace{-5pt}\fi
\end{table}

Portions of the term table, formula table, and string table for our example proof from Section~\ref{sec:redos} appear in Tables~\ref{tab:expr-table-from-example}, \ref{tab:formula-table}, and~\ref{tab:string-table}, respectively.
In our tables for AST nodes, the field $\nodeid$ indicates the specific constructor used in our grammar for an entry.
In $\termTable$, an entry's $\nodeid$ denotes the kind of term node that it represents.
In $\formulaTable$, the $\nodeid$ is the predicate that a formula uses.
Entries in $\syncTable$ have no $\nodeid$ because they are all strings.
Also, entries in $\termTable$ and $\syncTable$ have a field $\varid$ that represents the character stored at a node.
$\varid$ is an immediate value rather than a pointer to some other location.
For all three AST tables, pointers are stored in the field $\arglist$, regardless of whether they point to other entries in the same table or to entries in different tables.

Along with the three tables for AST nodes,
there is a fourth read-only table, the \textit{step table} $\stepTable$.
Table~\ref{tab:intro-proof} from Section~\ref{sec:redos} is an example of a step table.
Each entry in $\stepTable$ represents a step in the proof.
The steps in $\stepTable$ do not necessarily appear in their underlying logical order.
A step table entry includes an index indicating its logical order within the proof, an ID for the rule it uses, an ID for that rule's multiplexing category (Section~\ref{sec:zk implementation}), a pointer to the step's conclusion in $\formulaTable$, and pointers to the step's premises in $\formulaTable$.
The formula table and the step table always have the same number of entries.
We keep the two tables distinct to make ZK execution of \tool more efficient (Section~\ref{sec:zk implementation}).

\ifsubmission
\else
\begin{table}
\centering
\begin{tabular}{|@{\hskip 3pt}c@{\hskip 3pt}|@{\hskip 3pt}c@{\hskip 3pt}|@{\hskip 3pt}l@{\hskip 3pt}|@{\hskip 3pt}c@{\hskip 3pt}|}
    \hline 
    {\sf\bf Addr.} & {\bf\nodeid} & {\bf\arglist}  & {\bf\sf Meaning}\\
    \hline
    $\forad{0}$ & \eq &  $\{\exad{10},\exad{3} \}$ & $\derivative a{a^*}=a^*$ \\
    \hline
    $\forad{1}$ & \eq & $\{\exad{6},\exad{1} \}$ & $\hasblank{a^*}=\regexBlank$ \\
    \hline
    $\forad{2}$ & \eq & $\{\exad{5},\exad{3} \}$ & $(a^*a)^*=a^*$ \\
    \hline
    $\forad{3}$ & \syncNode & $\{\chrad{1},\exad{12},\exad{3} \}$ & $\Sync{a}{(a^*a\regexUnion\regexBlank)(a^*a)^*}{a^*}$ \\
    \hline
    $\forad{4}$ & \syncNode & $\{\chrad{0},\exad{12},\exad{3} \}$ & $\Sync{\regexBlank}{(a^*a\regexUnion\regexBlank)(a^*a)^*}{a^*}$ \\
    \hline
    \end{tabular}
    \captionof{table}{Part of the formula table $\formulaTable$ for the proof of the equivalence of $(a^*a)^*$ and $a^*$. We use $\forad{}$ to denote the addresses of formulas.}
    \label{tab:formula-table}
\end{table}
\fi

In the proof in Table~\ref{tab:intro-proof}, different steps perform different functions depending on their rules.
Steps for unfolding derivatives and epsilon applications appear at the start of the proof.
The step labeled SyncCycle takes advantage of the fact that $(a^*a\regexUnion\regexBlank)(a^*a^*)$ and $a^*$ are both their own derivatives with respect to $a$ to begin the creation of a bisimulation.
The step labeled Coinduction continues the construction of the bisimulation, and the step labeled SyncEmpty converts the bisimulation into an equality predicate.
At the end, we conclude that $(a^*a)^*=a^*$.

\begin{table}
\centering
\begin{tabular}{|@{\hskip 2pt}c@{\hskip 2pt}|@{\hskip 2pt}c@{\hskip 2pt}|@{\hskip 2pt}c@{\hskip 2pt}|@{\hskip 2pt}l@{\hskip 2pt}|}
    \hline 
    {\sf\bf Addr.} & {\bf\varid } & {\bf\arglist}  & {\bf\sf Meaning}\\
    \hline
    $\chrad{0}$  & $\regexBlank$ & $\{\}$ & $\regexBlank$ \\
    \hline
    $\chrad{1}$  & $a$ & $\{\chrad{0}\}$ & $a$ \\
    \hline
    $\chrad{2}$  & $b$ & $\{\chrad{1}\}$ & $ba$ \\
    \hline
    \end{tabular}
    \captionof{table}{The string table for the proof of the equivalence of $(a^*a)^*$ and $a^*$.  We use $\chrad{}$ to denote the addresses of strings.  We include an extra row to show how we represent multi-character strings.}
    \label{tab:string-table}
\end{table}

Proofs in \tool's calculus start with a single assumption,
but apart from that,
there are no contexts, environments, or temporary assumptions for proof steps.
Whenever we derive a conclusion, we know that it holds unconditionally under the starting assumption.

\paragraph{Checking Instructions}

\tool relies on a calculus of more than forty distinct proof rules.
Every proof rule has its own corresponding \textit{checking instruction}.
A checking instruction is a function that confirms that an individual application of a specific proof rule is valid.
Most checking instructions only perform simple pattern matching on the premises and conclusion of a proof step.
For instance, the checking instruction for Trans (transitivity of equality) starts by fetching the step's premises $\formula_1$ and $\formula_2$ and its conclusion $\formula_0$ from $\formulaTable$.
It asserts that all three formulas have $\eq$ as their $\nodeid$.
Next, the checking instruction asserts some pointer equalities: $\formula_1.\arglist[0]=\formula_0.\arglist[0]$, $\formula_1.\arglist[1]=\formula_2.\arglist[0]$, and $\formula_2.\arglist[1]=\formula_0.\arglist[1]$.
These assertions suffice to confirm that a proof step derives a conclusion of the form $x=z$ from premises of the form $x=y$ and $y=z$.
Importantly, the checking instruction never needs to fetch $x$, $y$, or $z$ from $\termTable$.
To confirm that two occurrences of $x$, $y$, or $z$ are equal, it only needs to check that the same pointer appears in both formulas.
The $\arglist$ entries in all three formulas are pointers to the term table for $x$, $y$, and $z$.
In general, when we need to check for equality between two terms or strings, we only check pointer equality.
Sometimes we perform more complex linear-time scans on strings or chains of derivatives, but not for mere equality checking.
In Appendix~\ref{sec:instructions_in_zk}, we discuss the more complex checking instructions in more detail.

\paragraph{Size Parameters}

We use five numerical size parameters and one set to define the characteristics of an instance of \tool.  All of them are public, and each one relates to a different component of the proof.
\textbf{(1)} $\alphabetsize$ is the number of distinct characters in the alphabet $\alphabet$.
\textbf{(2)} $\termTableLength$ is the number of entries in $\termTable$.
\textbf{(3)} $\syncTableLength$ is the number of entries in $\syncTable$.
\textbf{(4)} $\stepTableLength$ is the number of entries in $\formulaTable$.
\textbf{(5)} $\maxSyncLength$ is the maximum length of any individual string stored in $\syncTable$.
Lastly, \textbf{(6)} $\theories$ is a set of sets of checking instructions that \tool can use.
Every set $\theory$ within $\theories$ is a distinct category for multiplexing.
We discuss multiplexing in more detail in Section~\ref{sec:zk implementation}.

\paragraph{Execution}

Algorithm~\ref{alg:exe} shows how \tool validates equivalence proofs.
\tool iterates over the entries in the step table, checking that each one is valid.
At the end, we know that the whole proof is valid because every individual step is valid.

\SetKwFunction{KwSubchecker}{{CheckingInstrs}}
\SetKwFunction{KwPermchecker}{PermuteCheck}
\SetKwFunction{KwConsistency}{ConsistencyCheck}
\begin{algorithm}[!t]
\caption{$\tool[\theories](\alphabetsize, \termTableLength, \syncTableLength, \stepTableLength, \maxSyncLength)$}\label{alg:exe}

$\validsteps \leftarrow [0, \dots, 0]$\;

\For { ${\sf pc}=0$ to $\stepTableLength-1$ \label{line:iter-pf-steps}}{
\!{\bf Proof Step Fetch: \label{line:pf-fetch-label}}\\
$\proofstep_0\leftarrow \stepTable[{\sf pc}]$\;\label{alg:bind-start}
$\theory\leftarrow\theories[\proofstep_0.\categoryid]$\;\label{line:get-category}
\!{\bf Conclusion Fetch:}\\
$\formula_0 = \formulaTable[\proofstep_0.\resultShort]$\;\label{alg:bind-concl}

\!{\bf Premise Fetch:}\\
${\proofstep_1, \proofstep_2} \leftarrow  \stepTable[\proofstep_0.\premisesShort[0]], \stepTable[\proofstep_0.\premisesShort[1]]$\;\label{alg:bind-premises}
${\formula_1, \formula_2} \leftarrow  \formulaTable[\proofstep_1.\resultShort], \formulaTable[\proofstep_2.\resultShort]$\;\label{alg:bind-premise-formulas}
\!{\bf Rule Checking:} \label{line:rule-chk-label} \\
$z\leftarrow\text{false}$\;
\For{$\proofrule\in\theory$}{
$z\leftarrow z\lor\KwSubchecker[\proofrule](\formula_0, \{\formula_1, \formula_2\})$\;\label{alg:chking-instr}
}
${\bf assert}(z)$\;\label{line:multiplex}
\!{\bf Cycle Checking:} \label{line:cycle-chk-label} \\
${\bf assert}(\proofstep_1.\stepid < {\proofstep_0.\stepid})$\; \label{alg:assert-acyclic-proof}
${\bf assert}(\proofstep_2.\stepid < {\proofstep_0.\stepid})$\;\label{line:cycle-end}
$\validsteps[{\sf pc}] \leftarrow {\proofstep_0.\stepid}$\; \label{line:store-d}
}
$\KwPermchecker(\validsteps, [0, \dots, \stepTableLength -1])$\;\label{alg: permute}
$\KwConsistency(\termTable,\syncTable)$\;\label{alg: type}
\end{algorithm}

\tool confirms that proof steps are valid by applying checking instructions to them.
Rather than applying only one checking instruction to each step, \tool multiplexes over a set of checking instructions (lines~\ref{line:get-category} and~\ref{line:rule-chk-label}--\ref{line:multiplex}).
Only one of the checking instructions needs to return a positive result for a given proof rule.
Multiplexing allows us to hide the specific proof rule used for a step, which becomes important when we validate proofs in ZK (Section~\ref{sec:zk implementation}).

There is no need to perform type checking for our table of terms because all terms have the same type.
Likewise, there is no need to perform type checking for our table of formulas $\formulaTable$ or our table of strings $\syncTable$.
However, we do need to perform some consistency checks on the tables apart from the checks performed by checking instructions.
We confirm that there are no more than $\alphabetsize$ distinct characters, and we confirm that there are no cyclic pointers in any of the tables (lines~\ref{line:cycle-chk-label}--\ref{line:cycle-end} and~\ref{alg: type}).
Also, at the end, we confirm that every proof step has been checked once (line~\ref{alg: permute}).

\section{Proof Rules}
\label{sec:proof-rules}

We will now explain the calculus of rules that \tool employs in its proofs.
Most of the rules are based on existing work, but we introduce some custom rules for coinduction.
Our proof rules are sound and complete for regular expression equivalence.
Our coinduction rules are the only ones whose soundness is non-trivial to establish, and we confirm their soundness as we introduce them.
Our completeness proof appears in Appendix~\ref{sec:complete}.

\paragraph{Rule Design}
We designed all of our proof rules to be checkable in either constant time or linear time relative to $\maxSyncLength$.
The proof rules for ZKSMT~\cite{luick2023zksmt} and zkPi~\cite{cryptoeprint:2024/267} follow a similar pattern.
Keeping all of our proof rules simple allows us to check them easily in ZK without applying excessive padding.
If we had individual proof steps that traversed ASTs of arbitrary depth, we would need to apply a significant amount of padding to hide the size and shape of the AST being traversed.
Every application of a tree-traversing rule would need to incur the same cost, and that cost would scale relative to the maximum AST size.

For a similar reason, each proof rule takes at most two premises.
In Table~\ref{tab:unified} we depict the rules Coinduction and Match as taking $\alphabetsize+1$ premises, but, in the underlying implementation, we split them into multiple rules, none of which take more than two premises.

\paragraph{Simple Proof Rules}
Our rules other than the coinduction rules fall into five main categories.
\textbf{(1)} Our \textit{epsilon rules} unfold the definition of $\hasblankFunction$.
Instead of traversing a whole AST, each rule unfolds the definition of the epsilon function for one AST node.
They follow the standard definition of the epsilon function.
\textbf{(2)} Our \textit{derivative rules} unfold the definition of $\derivativeFunction$.
They follow the standard definition of the derivative function, and their design is similar to the design of the epsilon rules.
\textbf{(3)} Our \textit{equality rules} are standard axioms of first-order logic.
They allow us to use equality as an equivalence relation and to perform substitutions of equivalent terms.
\textbf{(4)} Our \textit{normalization rules} are standard axioms of Kleene algebra for manipulating unions and concatenations~\cite{kozen1997kleene}.
Kleene algebra is the mathematical formalism underlying regular expressions.
Within our algorithm for proof generation, we use the normalization rules to convert regular expressions into normal form (Section~\ref{sec:proof-gen}, Algorithm~\ref{alg:proof-gen}).
\textbf{(5)} Our \textit{proof completion rules} bookend our proofs by introducing an assumption that two regular expressions are not equivalent at the start and deriving $\bot$ from it at the end.
A more detailed explanation of our simple rules appears in Appendix~\ref{sec:rules-extra}.

\subsection{Coinduction Rules}
\label{sec:coinduction-rules}

Our coinduction rules appear in Table~\ref{tab:unified}.
Before we explain our coinduction rules, we need to explain the meaning of the Sync predicate that they utilize.
The formal meaning of Sync depends on the concept of \textit{reducible counterexamples}.
Let $w$ be a string on which the terms $p$ and $q$ disagree.
We say that $w$ is a reducible counterexample if it can be expressed in the form $w=stu$, where $t$ is non-empty, such that $\derivative sp=\derivative{st}p$ and $\derivative sq=\derivative{st}q$.
If $w=stu$ is reducible, then $p$ and $q$ also disagree on $su$, which is a strictly shorter string.
Note that it is possible for $su$ to be reducible as well.
Also, if $p=\derivative wp$ and $q=\derivative wq$, then we can define $t$ as $w$ and have $s$ and $u$ be empty.
An irreducible counterexample for $p$ and $q$ is simply a counterexample that is not reducible.
Formally, $\Sync spq$ means that $p$ and $q$ do not have any irreducible counterexamples that start with $s$.

\begin{table}[!t]
    \centering
    \begin{tabular}{|c|c|c|}
    \hline 
    {\sf RuleID} & {\premise} & {\sf Conclusion}\\    
    \hline

    Match
    &
    $\hasblank p=\hasblank q$,
    &
    $p=q$
    \\
    &
    $\forall c. \derivative cp=\derivative cq$
    &
    \\
    \hline

    Coinduction
    &
    $\hasblank{\derivative sp}=\hasblank{\derivative sq}$,
    &
    $\Sync{s}pq$
    \\
    &
    $\forall c.\Sync{sc}pq$
    &
    \\
    \hline

    SyncCycle
    &
    $\derivative{cs}p=p$, $\derivative{cs}q=q$
    &
    $\Sync{cs}pq$
    \\
    \hline
    
    SyncFold
    &
    $\Sync s{\derivative cp}{\derivative cq}$
    &
    $\Sync{cs}pq$
    \\
    \hline

    EqualSync
    &
    $\derivative sp=\derivative sq$
    &
    $\Sync spq$
    \\
    \hline

    SyncEmpty
    &
    $\Sync\regexBlank pq$
    &
    $p=q$
    \\
    \hline
    
    \end{tabular}
    \caption{Coinduction Rules}
    \label{tab:unified}
\end{table}

\paragraph{Main Coinduction Rules}

Match and Coinduction
are the two most important rules in our calculus.
Match is simply the application of a key observation from prior work:
if $p$ and $q$ are two regular expressions with the same alphabet $\alphabet$, then $p$ and $q$ are equivalent if and only if $\hasblank p =\hasblank q$ and $\derivative cp=\derivative cq$ for every $c\in\alphabet$~\cite{antimirov1995rewriting}.
Match takes $\alphabetsize+1$ premises, each of which corresponds to a different part of the observation.
The premise $\hasblank p =\hasblank q$ means that $p$ and $q$ agree on the empty string.
Also, if $\derivative cp=\derivative cq$ for some $c$, then $p$ and $q$ agree on all strings that start with $c$.
If $p$ and $q$ are not equivalent, then there must exist some string that one accepts and the other rejects.
The premises $\hasblank p =\hasblank q$ and $\forall c\in\alphabet . \derivative cp=\derivative cq$ eliminate the possibility of $p$ and $q$ disagreeing on any string, so $p$ and $q$ must be equivalent.

Coinduction also takes $\alphabetsize+1$ premises.
The first of its premises, $\hasblank{\derivative sp}=\hasblank{\derivative sq}$,
establishes that $p$ and $q$ agree on $s$.
For the other $\alphabetsize$ premises, we have $\Sync{sc}pq$ for every $c\in\alphabet$.
Each of those premises gives us that $p$ and $q$ have no irreducible counterexamples that start with $sc$.
We know from the first premise that $s$ itself is not a counterexample for $p$ and $q$, so putting all of the premises together gives us that
$p$ and $q$ have no irreducible counterexamples that start with $s$.
This is precisely the meaning of $\Sync spq$, which is the conclusion of Coinduction.

\paragraph{Auxiliary Coinduction Rules}

Along with Match and Coinduction, we have some additional rules for manipulating Sync formulas.
The rule SyncCycle takes advantage of cycles in terms' derivatives.
If $\derivative{cs}p=p$ and $\derivative{cs}q=q$ for a character $c$ and string $s$, then we know that $\Sync{cs}pq$ holds.
We can say in this situation that, if $p$ and $q$ have a counterexample that starts with $cs$, that counterexample is reducible.
Suppose that $p$ and $q$ disagree on $cst$ for some string $t$.
We can express $cst$ as $s't'u$, where $s'=\regexBlank$ and $t'=cs$.
With this rephrasing, our premises become $\derivative{s't'}p=\derivative{s'}p$ and $\derivative{s't'}q=\derivative{s'}q$.
The string $t'=cs$ is non-empty, so these are precisely the requirements for reducibility, and $p$ and $q$ have no irreducible counterexamples that start with $cs$.
In other words, $\Sync{cs}pq$ holds, so SyncCycle is sound.

Importantly, SyncCycle cannot be applied with an empty string.
Every term is its own derivative with respect to $\regexBlank$.
If we did not enforce non-emptiness, we could derive $\Sync{\regexBlank}pq$ trivially for any $p$ and $q$ and then apply SyncEmpty to derive $p=q$, which would be unsound.

The rule SyncFold concludes $\Sync{cs}pq$ from the premise $\Sync s{\derivative cp}{\derivative cq}$.
The soundness of SyncFold follows from the definition of the derivative:
$\derivative cp$ accepts the set of strings $s$ such that $p$ accepts $cs$, and $\derivative cq$ accepts the set of strings $s$ such that $q$ accepts $cs$.
The conclusion $\Sync{cs}pq$ means that $p$ and $q$ have no irreducible counterexamples that start with $cs$.
Suppose that $p$ and $q$ do have an irreducible counterexample $cst$ for some string $t$.
Since $p$ and $q$ disagree on $cst$, it must hold that $\derivative cp$ and $\derivative cq$ disagree on $st$.
Moreover, $st$ must be an irreducible counterexample for $\derivative cp$ and $\derivative cq$.
If $st$ can be expressed as $s't'u$ such that $\derivative{s'}{\derivative cp}=\derivative{s't'}{\derivative cp}$ and $\derivative{s'}{\derivative cq}=\derivative{s't'}{\derivative cq}$, then that would violate our assumption that $cst$ is irreducible because $cst=cs't'u$, $\derivative{cs'}{p}=\derivative{cs't'}{p}$, and $\derivative{cs'}{q}=\derivative{cs't'}{q}$.
The fact that $\derivative cp$ and $\derivative cq$ have an irreducible counterexample that starts with $s$ contradicts $\Sync s{\derivative cp}{\derivative cq}$, our premise from the start.
Therefore, $p$ and $q$ cannot have any irreducible counterexamples that start with $cs$, and SyncFold is sound.

The rule EqualSync allows us to conclude $\Sync spq$ from $\derivative sp=\derivative sq$.
To see why EqualSync is sound, consider the string $st$ for some $t$.
By the definition of the derivative, $p$ accepts $st$ if and only if $\derivative sp$ accepts $t$.
Likewise, $q$ accepts $st$ if and only if $\derivative sq$ accepts $t$.
Therefore, $p$ and $q$ agree on $st$ if and only if $\derivative sp$ and $\derivative sq$ agree on $t$.
Our premise $\derivative sp=\derivative sq$ gives us that $\derivative sp$ and $\derivative sq$ agree on all strings, so $p$ and $q$ must agree on all strings that start with $s$.
If $p$ and $q$ have no counterexamples that start with $s$, they have no irreducible counterexamples that start with $s$, so $\Sync spq$ holds, and EqualSync is sound.

The rule SyncEmpty takes $\Sync{\regexBlank}pq$ as a premise and derives $p=q$.
The premise $\Sync\regexBlank pq$ means that $p$ and $q$ have no irreducible counterexamples that start with $\regexBlank$.
All strings start with $\regexBlank$, so
$p$ and $q$ have no irreducible counterexamples at all.
This means that $p$ and $q$ must be equivalent:  two regular expressions cannot have a counterexample unless they have an irreducible counterexample.
By definition, every reducible counterexample has a strictly shorter counterexample that can be constructed from it.
All strings are finite, so it cannot be the case that all counterexamples for a pair of inequivalent regular expressions are reducible.
Since $p$ and $q$ do not have any counterexamples, they must be equivalent, which is what we wanted to show.

\section{Zero-Knowledge Encoding}
\label{sec:zk implementation}

On its own, the basic structure of an instance of \tool does not hide the regular expressions being compared or the proof of their equivalence.
We need to use multiple zero-knowledge proof techniques to hide the structure of our proofs.
We use a commit-and-prove ZK protocol~\cite{canetti2002universally} to hide the terms and formulas manipulated by a proof, we apply padding when validating certain proof steps, and we multiplex over different checking instructions.
\tool is not necessarily tied to any specific cryptographic backend.
Our implementation uses many of the same ZK operations as ZKSMT~\cite{luick2023zksmt}, including
VOLE-ZK commitment~\cite{weng2021wolverine,baum2021mac,ITC:DitIshOst20},
optimizations for polynomials~\cite{yang2021quicksilver},
and oblivious RAM~\cite{franzese2021constant}.
There are resemblances to the ZK techniques used by zkPi~\cite{cryptoeprint:2024/267} as well.

\paragraph{Table Commitment}

Most of the proof structure in an instance of \tool appears in the tables $\termTable$, $\syncTable$, $\formulaTable$, and $\stepTable$, so we commit the entries of the tables.
We can represent any term, string, formula, or proof step as a vector of bits because the entries in our tables are all tuples of numbers, where those numbers either represent values on their own or serve as pointers to other table entries.
We can commit every bit in a table entry individually and then combine the committed bits for an entry into a single committed integer~\cite{franzese2021constant}.
When we commit the rows in our tables, the verifier can check equality between table entries and perform arithmetic operations on them without learning their values.

To prevent information leakage, all rows in an individual table are padded to be the same size.
Every row in a table has the maximum number of entries for its type, even if some of those entries are unused.
For instance, not all terms have two children, and not all formulas have pointers to $\syncTable$, but, in our ZK proofs, the nodes that do not have those fields have padding in their place.
There is no need to pad rows with fields for other types (such as pointers to $\syncTable$ in the term table) because all entries in a table are of the same type.

Additionally, we use oblivious indexing for $\termTable$, $\syncTable$, and $\formulaTable$.
\tool does not modify its tables during execution, so we can use a ZK protocol for read-only memory for the tables~\cite{heath20202,franzese2021constant,SCN:DOTV22}.
With oblivious indexing, the verifier can use a committed integer index to retrieve a row of a table without learning the value of the index or the entry that was retrieved.
This means that the verifier cannot perform frequency analysis on the memory fetches that occur during proof validation to recover extra information.

Although we commit the rows in $\stepTable$, we do not support oblivious indexing for $\stepTable$.
During proof validation, every entry of $\stepTable$ is fetched exactly once, so there is no need to safeguard against frequency analysis for the step table.

The amortized time cost of fetching an entry from ZK memory is linear in the bit width of the entries and does not depend on the number of rows in the memory.
Our memory entries are all of constant size:  they do not vary with the size parameters for a proof.
Consequently, memory fetches are effectively a constant-time operation for \tool.
ZKSMT's read-only memory has the same amortized performance~\cite{luick2023zksmt}.

\paragraph{Multiset Operations}

Unlike ZKSMT, \tool does not have a dedicated table for storing lists.
Nevertheless, the protocol needs to reason about lists and multisets during validation.
For multisets, the order of elements is irrelevant, but their multiplicities matter.
Whenever we need to confirm that two lists are equivalent when viewed as multisets, we convert the lists into polynomials over a finite field and use a check similar to the checks used by ZKSMT~\cite{luick2023zksmt}.

\paragraph{Multiplexing}

Just like ZKSMT~\cite{luick2023zksmt}, \tool shuffles the steps in its proofs to hide their logical ordering.
As a further guard against information leakage, \tool uses \textit{multiplexing}:  when it validates a proof step, it runs multiple checking instructions on the step to hide the rule used for the step.

The simplest and most secure approach for multiplexing would be to run every checking instruction on every step.
However, in Section~\ref{sec:eval-validation}, we observe that this approach causes \tool to run very slowly on large proofs.
Instead, we group the rules in our calculus into three main categories for multiplexing.
If the rule for a proof step $\proofstep$ is in the category $\theory$, then \tool runs every checking instruction in $\theory$ on $\proofstep$.
Consequently, when
\tool runs on a proof, it leaks the number of proof steps in each multiplexing category.
This is a significant improvement over the information leakage of ZKSMT, which leaks the frequencies of all distinct proof rules~\cite{luick2023zksmt}.

The checking instructions that run in linear time relative to $\maxSyncLength$ have their own multiplexing category.
If we placed linear-time rules in the same category as everything else, every proof step would take linear time to check.
Also, we split the constant-time checking instructions into two categories:  one for Symm, Trans, and FunCong2 (Appendix~\ref{sec:rules-extra}, Table~\ref{tab:equality}), and the other for everything else.
We have a separate category for Symm, Trans, and FunCong2 because they are consistently the most commonly used rules in the regular expression equivalence proofs that our proof generation algorithm generates.
The division of constant-time rules into two categories is an arbitrary choice for performance optimization, not a requirement for avoiding information leakage.
It is similar to the approach taken by zkPi~\cite{cryptoeprint:2024/267}, which separates judgments into two categories for multiplexing for the sake of efficiency.
A more detailed discussion of the ZK checking instructions that require linear time appears in Appendix~\ref{sec:instructions_in_zk}.

\paragraph{Proof Step Ordering}
Like ZKSMT~\cite{luick2023zksmt}, \tool performs a permutation check to ensure that every proof step has been checked once.
Every proof step has a committed ID, and a proof step cannot take other steps as premises unless those steps' IDs are strictly lower than its own (Algorithm~\ref{alg:exe}, lines~\ref{alg:assert-acyclic-proof}--\ref{line:cycle-end}).
Every time \tool validates a proof step, it adds that step's ID to the array $\validsteps$.
At the end of validation, \tool confirms that every proof step's ID appears in $\validsteps$ exactly once.
We can confirm this in linear time by performing a multiset equivalence check between $\validsteps$ and the list of all step IDs from 0 to $\stepTableLength-1$ (Algorithm~\ref{alg:exe}, line~\ref{alg: permute}).
This technique for checking that one list is a permutation of another comes from~\cite{blum1994checking} originally, and ZKSMT uses it as well~\cite{luick2023zksmt}.

\paragraph{Information Leakage}

All of \tool's size parameters are public.
Along with the size parameters, \tool leaks the number of proof steps in each multiplexing category,
but it does not leak the specific rule used for each step.
In contrast, ZKSMT leaks the number of uses of each proof rule~\cite{luick2023zksmt}.

The fact that we store terms, strings, and formulas in three different tables is not a source of information leakage.
Even if we stored all three in the same table, the verifier would always know whether an entry being fetched at a given point during execution is a term, string, or formula since the verifier knows the definitions of the checking instructions.

On its own, \tool does not require the prover to reveal the regular expressions being proven equivalent.
However, the prover can selectively leak information about one or both of them to ground the proof in a larger system.
Depending on the use case, the verifier can demand information about the starting regular expressions to prevent the prover from providing an irrelevant proof.

\paragraph{Knowledge Soundness and Zero-Knowledge}
The fact that \tool upholds knowledge soundness and zero-knowledge follows immediately from the fact that its cryptographic backend does~\cite{weng2021wolverine}.
The verifier validates every ZK operation in a proof individually,
so there are no opportunities for the prover to falsify a ZK operation or for the verifier to learn anything about the ZK operations other than the number of operations performed.
If we multiplex over all checking instructions for every step (Section~\ref{sec:eval-validation}), the verifier gains no information from the relative ordering of the ZK operations:
the size parameters
$\alphabetsize$, $\termTableLength$, $\syncTableLength$, $\stepTableLength$, and $\maxSyncLength$
are the only information leaked.
Also, the prover can always make the size parameters larger than they need to be to reduce information leakage even further.
This is analogous to the information leakage of
ZKUNSAT~\cite{luo2022proving} and zkPi~\cite{cryptoeprint:2024/267}.

\section{Proof Generation Process}
\label{sec:proof-gen}

Now we will explain our process for generating proofs that \tool can validate.
We use two algorithms:  the first checks whether two regular expressions are equivalent, and the second generates an equivalence proof for two equivalent regular expressions.

\subsection{Equivalence Checking}
\label{sec:equiv-check}

\begin{algorithm}[!t]
\caption{Equivalence check $\equivAlg(H,p,q)$}\label{alg:search}

$p'\leftarrow\normalize{p}$\;

$q'\leftarrow\normalize{q}$\;

\If{$p'\identical q'$}{return true\;}
\If{$(p',q')\in H$}{return true\;}
$e_1\leftarrow\normalize{\reduce{\hasblank{p'}}}$\;
$e_2\leftarrow\normalize{\reduce{\hasblank{q'}}}$\;
\If{\textbf{not }{($e_1\identical e_2$)}}{return false\;}

$v\leftarrow\text{true}$\;

\For{$c\in\Sigma$}{
$p_c\leftarrow\reduce{\derivative{c}{p'}}$\;
$q_c\leftarrow\reduce{\derivative{c}{q'}}$\;
$v\leftarrow v\land \equivAlg(H\cup\{(p',q')\},p_c,q_c)$\;
}
$\text{return }v$\;
\end{algorithm}

Algorithm~\ref{alg:search} checks whether two regular expressions have the same language.
$\equivAlg$ is a reformulation of an algorithm from~\cite{almeida2009testing} in our syntax.
The algorithm from~\cite{almeida2009testing} is itself based on
decision procedures from~\cite{antimirov1995rewriting} and~\cite{almeida2008antimirov}.
At a high level, $\equivAlg$ is simply an exhaustive search over the derivatives of the two regular expressions being compared.
It terminates its exploration of a path when it finds a cycle or a case where $p$ and $q$ disagree.
$\equivAlg$ uses the accumulator set $H$ to detect cycles:  $H$ stores the derivative pairs that it has encountered previously.
Different recursive paths have their own versions of $H$, so derivative pairs are flagged as repeats only if they appeared previously in the path currently being explored.
At the start of execution, $H$ should be empty.

$\equivAlg$ is guaranteed to terminate because of a result from prior work~\cite{brzozowski1964derivatives}.
For a regular expression $r$, the set of all unfolded derivatives $\reduce{\derivative sr}$ for strings $s$ can be infinite, but those derivatives must fall into a finite number of equivalence classes for similarity.
Similarity is an equivalence relation:  two regular expressions are similar if one can be converted into the other using our normalization and equality rules (Appendix~\ref{sec:rules-extra}).
The operator $\identical$ checks whether two terms are syntactically identical.
The function $\reduceFunction$ unfolds all derivative and epsilon nodes in a term using our proof rules.
The function $\normalizeFunction$ returns a normalized version of a regular expression:  it maps all regular expressions in a similarity class to the same unique normal form.
Our version of similarity is more general than the version in~\cite{brzozowski1964derivatives}, but making similarity more inclusive preserves the result.
Because the sets of similarity classes for the derivatives of $p$ and $q$ are finite, the recursion in Algorithm~\ref{alg:search} is guaranteed to find a cycle with $H$ eventually on any path it takes.

\subsection{Proof Generation}
\label{sec:proof-gen-alg}

To generate proofs of two regular expressions' equivalence rather than simply finding a Boolean result, we extend Algorithm~\ref{alg:search}.
Algorithm~\ref{alg:proof-gen} follows the same strategy as $\equivAlg$, but it records its path exploration in the form of a proof.
$\proofAlg$ does not check on its own that $p$ and $q$ are equivalent, but it produces a valid proof if they are.
It has an accumulator set $H$ that serves the same purpose as the one in $\equivAlg$.
Additionally, it uses an accumulator string $s$ that should be $\regexBlank$ initially.
The string records all of the derivative characters used so far to reach the current point in the algorithm's execution, and we use that information for our coinduction rules.

\begin{algorithm}[!t]
\caption{Proof generation $\proofAlg(H,s,p,q)$}\label{alg:proof-gen}

$(p',\proofstep_p)\leftarrow\normalize{p}$\;

$(q',\proofstep_q)\leftarrow\normalize{q}$\;

\If{$p'\identical q'$}{
$\proofstep_r\leftarrow\text{Refl}(p',q')$\;
$\proofstep_r'\leftarrow\text{Subst}(\proofstep_r,\proofstep_p,\proofstep_q)$\;
return $\text{EqualSync}(\proofstep_r')$\;
}
\If{$\exists s_0.(p',q',s_0)\in H$}{
$\proofstep_h\leftarrow\text{SyncCycle}(\texttt{suffix}(s_0,s),p',q')$\;
return $\text{Subst}(\proofstep_h,\proofstep_p,\proofstep_q)$\;
}
$(e_1,\proofstep_1)\leftarrow\normalize{\reduce{\hasblank{p'}}}$\;
$(e_2,\proofstep_2)\leftarrow\normalize{\reduce{\hasblank{q'}}}$\;
$\proofstep_e\leftarrow\text{Refl}(e_1,e_2)$\;
$\proofstep_e'\leftarrow\text{Subst}(\proofstep_e,\proofstep_1,\proofstep_2)$\;

$\Psi\leftarrow[\proofstep_e']$\;

\For{$c\in\Sigma$}{
$(p_c,\proofstep_3)\leftarrow\reduce{\derivative{c}{p'}}$\;
$(q_c,\proofstep_4)\leftarrow\reduce{\derivative{c}{q'}}$\;
$\proofstep_c\leftarrow \proofAlg(H\cup\{(p',q',s)\},sc,p_c,q_c)$\;
$\Psi\leftarrow\Psi :: \proofstep_c$\;
}
$\proofstep_s=\text{Coinduction}(\Psi)$\;
return $\text{Subst}(\proofstep_s,\proofstep_p,\proofstep_q)$\;
\end{algorithm}

The versions of $\normalizeFunction$ and $\reduceFunction$ used in Algorithm~\ref{alg:proof-gen} return proofs that their main output is equivalent to their input.
The function $\texttt{suffix}(s_0,s)$ returns the characters at the end of $s$ that do not appear in $s_0$.
We know that $s_0$ must be a prefix of $s$, so the application of the function within Algorithm~\ref{alg:proof-gen} is always safe.
The function $\text{Subst}(\proofstep,\proofstep_1,\proofstep_2)$ creates a new proof tree where the equivalence proven by $\proofstep_1$ is used as a substitution on the left-hand side of $\proofstep$ and the equivalence proven by $\proofstep_2$ is used as a substitution on the right-hand side of $\proofstep$.
We can perform the substitution using our equality rules.
Lastly, Refl is one of our equality rules.

Algorithm~\ref{alg:proof-gen} is guaranteed to terminate if its two input regular expressions are equivalent.
If $p$ and $q$ are equivalent, Algorithm~\ref{alg:proof-gen} follows the same execution path as Algorithm~\ref{alg:search} in terms of recursion and branching.
In the underlying implementation of $\proofAlg$, we use some performance optimizations not shown in Algorithm~\ref{alg:proof-gen}.
For instance, we use Match instead of Coinduction whenever $\Psi$ contains only equality formulas and no Sync formulas.

\paragraph{Post-Processing}

Algorithm~\ref{alg:proof-gen} often generates distinct proof steps that have the same conclusion.
A proof that derives the same conclusion multiple times is redundant, so,
after the algorithm finishes, we perform a second pass to eliminate redundancies.
For any step $\proofstep$ with $\mu$ premises, let $\treesize{\proofstep}=1+\sum_{i=0}^{\mu-1}\treesize{\proofstep.\premisesShort[i]}$.
(If $\proofstep$ has no premises, then $\treesize{\proofstep}=1$.)
If multiple distinct proof steps $\proofstep_1,\dots,\proofstep_k$ within the proof returned by Algorithm~\ref{alg:proof-gen} have the same conclusion,
let $\hat\proofstep=\text{argmin}_{\proofstep\in\{\proofstep_1,\dots,\proofstep_k\}}\treesize{\proofstep}$.
Modify all of the proof steps that take $\proofstep_1,\dots,\proofstep_k$ as premises to take $\hat\proofstep$ as a premise instead.
This transformation preserves the correctness of the proof because the validity of a proof step depends only on the conclusions of its immediate premises, not on any other structural features of its premises.
Repeat this process until no redundant steps remain.

During the process of removing duplicates, it is possible for some steps to become disconnected from the main proof tree even if they are not duplicates.
Consequently, once we have eliminated all redundant steps, we discard any remaining steps that are not in the tree of steps leading to the final conclusion.
The end result is that we have a proof where every step has a distinct conclusion and no steps are unneeded.
We do not formally guarantee that the reduced proof has the smallest possible number of steps needed to derive its conclusion, but in practice our post-processing makes proofs much smaller than they would be otherwise.

\section{Evaluation}
\label{sec:eval}

In our evaluation, we seek to answer four research questions.
($\researchQ1$) Does $\proofAlg$ scale well relative to the size of the regular expressions it receives as input?
($\researchQ2$) Can \tool validate equivalence proofs for complex regular expressions?
($\researchQ3$) Does \tool scale well relative to increases in proof size?
($\researchQ4$) Does \tool run efficiently even when it uses multiplexing?

We report a positive answer for all four questions.
For our evaluation, we used AWS instances of type \texttt{r5b.4xlarge} with 128 GB of memory and 16 vCPUs.
For validation, we ran the prover and verifier on two separate instances with a 10 Gbps network connection between them.

\subsection{Benchmarks}
\label{sec:benchmarks}

Our regular expression benchmarks come from the evaluation suite of FlashRegex, a tool for generating regular expressions that are immune to denial-of-service attacks~\cite{li2020flashregex}.
Other existing regular expression benchmark suites, such as the suite used by ReDoSHunter~\cite{li2021redoshunter}, do not contain significant numbers of equivalent regular expression pairs.
The FlashRegex suite does not group regular expressions into equivalent pairs on its own, so we follow a two-phase process to generate proofs for our evaluation.
For the first phase, we use $\equivAlg$ to check whether two regular expressions are equivalent according to our formalization.
We use 1,475 distinct regular expressions from FlashRegex for a total of ${{1,475}\choose{2}}=1,087,075$ pairs to check for equivalence.
For the second phase, we run $\proofAlg$ on the equivalent pairs that we find with $\equivAlg$ to generate our proofs.

\begin{figure}[!t]
 \centering
\ifsubmission
\else\includegraphics[width=0.95\textwidth]{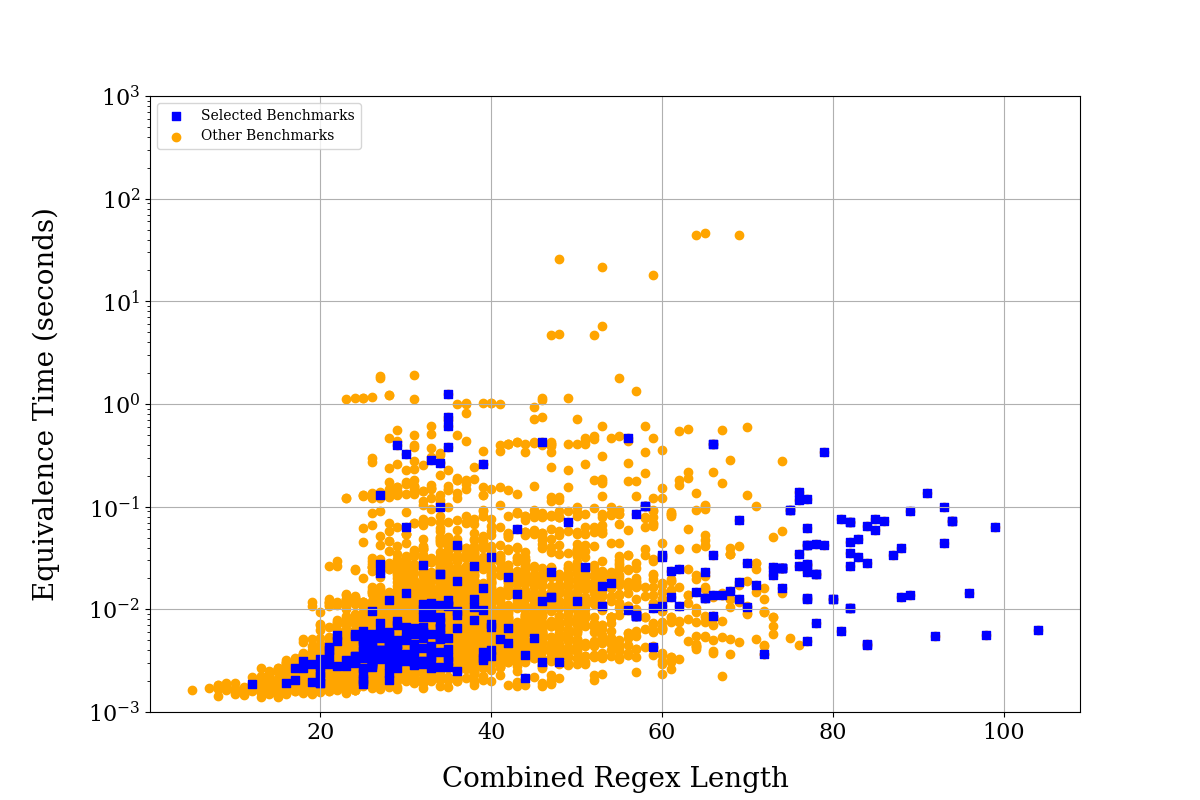}\fi
\caption{
Equivalence checking time relative to the combined size of the regular expressions being compared, plotted on a logarithmic scale.
}
\label{fig:equivcheck}
\end{figure}

Our time limit for $\equivAlg$ is ten minutes, and 21 pairs hit the time limit.
For a small number of inputs, $\equivAlg$ takes a long time to terminate, with the slowest taking 594.62 seconds, but the average running time is only 1.12 seconds for confirmed equivalent pairs and 0.0025 seconds for confirmed inequivalent pairs.
We plot the running times for $\equivAlg$ in Figure~\ref{fig:equivcheck}.
We also impose a ten-minute time limit for $\proofAlg$.
Of the 7,353 equivalent pairs that we find with $\equivAlg$, 72 hit the time limit during proof generation.
We exclude them from consideration for our evaluation.

Most of the regular expressions that we use for testing have two-character alphabets.
Regular expressions with small alphabets are more likely to have semantic equivalents than regular expressions with large alphabets are.
The benchmarks that hit the time limit for $\equivAlg$ and $\proofAlg$ involve regular expressions with high numbers of unions and stars.
All 21 timeouts for $\equivAlg$ and 13 of the 72 timeouts for $\proofAlg$ involve this regular expression:

\ifsubmission
\else
\begin{center}
\texttt{((aabb|abab|abba|bbaa|baba|baab)*(aa|ab|ba|bb)*)+((ab)?)}
\end{center}
\fi

Moreover, all 15 of the non-timeout inputs that cause $\equivAlg$ to run for more than 61 seconds involve the same regular expression.
Also, 33 of the $\proofAlg$ timeouts involve this regular expression:

\ifsubmission
\else
\begin{center}
\texttt{(bb|aa|(aabb|abab|abba|baba|bbaa|baab)*)*((aa)|(bb)|(ab)|(ba))*}
\end{center}
\fi

To keep our evaluation suite manageable, we select a sample of our proofs for validation instead of running \tool on all 7,281 of them.
Our main test suite includes the 50 proofs whose starting regular expressions have the highest combined length,
the 50 proofs with the highest value of $\syncTableLength$ (string table length),
the 50 proofs with the highest value of $\maxSyncLength$ (maximum Sync string length),
the 50 proofs with the highest value of $\stepTableLength$ (number of proof steps),
and 200 randomly selected proofs.
We have 301 main benchmarks in total rather than 400 because the categories overlap partially.

\subsection{Proof Generation Time}
\label{sec:proof-gen-eval}

\begin{figure}[!t]
 \centering
\ifsubmission
\else\includegraphics[width=0.95\textwidth]{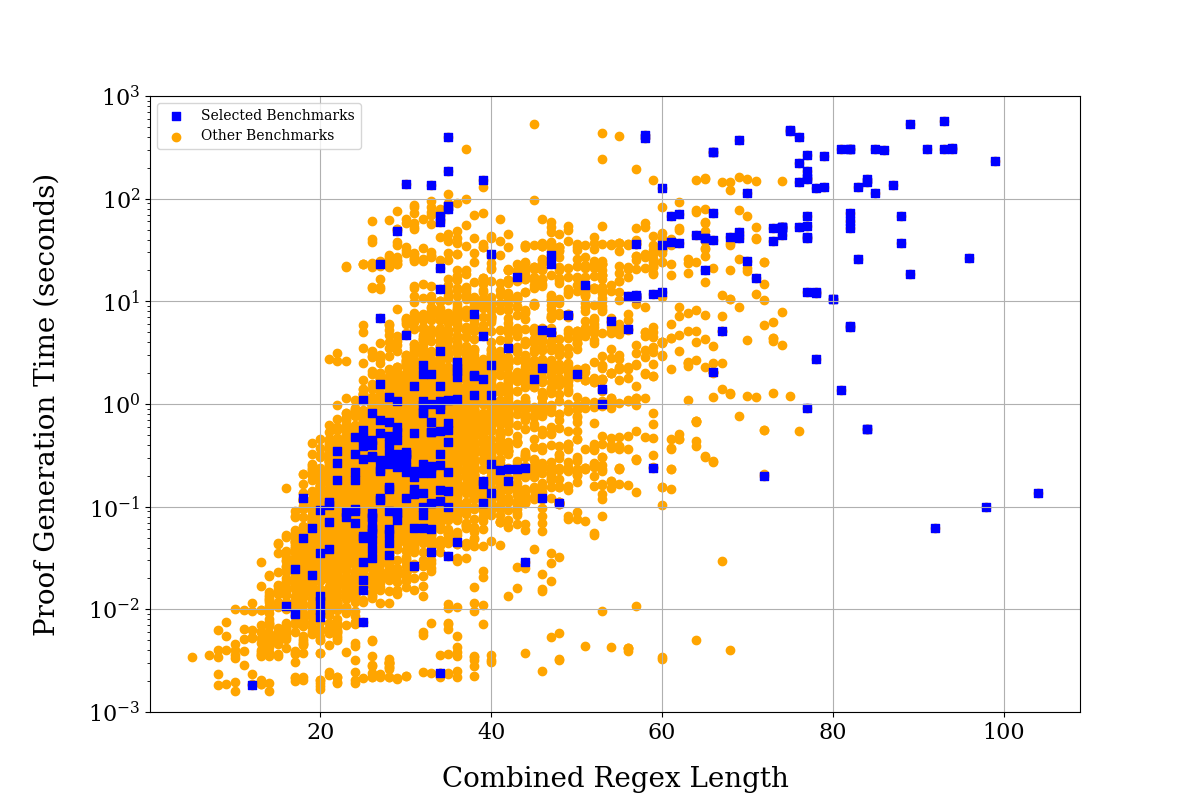}\fi
\caption{
Proof generation time relative to the combined size of the regular expressions being compared, plotted on a logarithmic scale.
}
\label{fig:proofgen}
\end{figure}

To answer $\researchQ1$,
we time the proof generation process for all equivalent pairs except the ones that hit the time limit for $\equivAlg$.
We plot the running time of our proof generator against the combined size in characters of the two regular expressions being compared.
The results appear in Figure~\ref{fig:proofgen}.
Our proof generation time includes the execution of $\proofAlg$ and the post-processing for the removal of redundant steps that we describe in Section~\ref{sec:proof-gen-alg}, but it does not include the execution of $\equivAlg$.

Across all 7,281 equivalent pairs for which proof generation succeeds, the average running time for proof generation is 4.66 seconds, the median is 0.29 seconds, and the standard deviation is 26.04 seconds.
There are a few very slow outliers, with the slowest taking 566.00 seconds, but proof generation is usually very quick.
We do not consider the slow outliers a problem.
Proof generation happens offline, before \tool runs.
It involves no cryptographic operations, and the verifier does not observe it.
Furthermore, because regular expression equivalence is PSPACE-complete, slow running times are inevitable for some large inputs.

The slowest non-timeout benchmark for $\proofAlg$ has one of the highest combined regex lengths,
but, in general, the correlation between combined regex length and proof generation time is not exact.
Across all the benchmarks that do not hit the time limit, the $R^2$ value for the correlation between the log of the running time of $\proofAlg$ and the combined regex size of the input is approximately 0.63.
The running time of $\proofAlg$ grows at a roughly exponential rate relative to the combined size in characters of the regular expressions being compared, but factors other than regex length can affect the running time of $\proofAlg$ significantly.
If a regular expression from the starting pair has an AST that is far removed from normal form, $\proofAlg$ will need to use a significant number of proof steps to normalize it.
Also, a regular expression's length and the complexity of its AST do not correlate perfectly.
For instance, parentheses do not appear in a regular expression's AST, but they still count toward its regex length.

We highlight the benchmarks that we select for validation in a different color in Figure~\ref{fig:proofgen}.
The proofs that we select for validation are generally larger and slower to generate than the rest of the proofs are.
Among the 301 equivalent pairs that we use for validation, the average running time for proof generation is 43.89 seconds, the median is 0.98 seconds, and the standard deviation is 98.88 seconds.
The maximum is 566.00 seconds, just as it is for the full suite.
For proof step counts, the benchmarks that we selected for validation have an average of 1839.82, a median of 890, and a standard deviation of 1853.98.
For comparison, the full collection of 7,281 equivalence proofs has an average proof step count of 807.79, a median of 652, and a standard deviation of 656.50.

\subsection{Comparison to Z3}

For a further assessment of $\researchQ1$, we performed a baseline comparison of $\proofAlg$ against Z3.
As we mentioned in Section~\ref{sec:regex-equivalence},
Z3 can check equivalences between regular expressions and generate proof certificates for them.
We ran Z3 on our 301 selected benchmarks with a time limit of ten minutes, both with and without proof generation enabled.
With proof generation enabled, Z3 hits the time limit for 8 of the benchmarks.
Among the 293 benchmarks for which Z3 does not hit the time limit, Z3 has an average running time of 0.088 seconds, a median of 0.070 seconds, and a standard deviation of 0.061 seconds.
Its slowest non-timeout running time is 0.42 seconds.
With proof generation disabled, Z3 is somewhat faster, but it hits the time limit on the same 8 benchmarks.
Among the 293 successes, it has an average running time of 0.061 seconds, a median of 0.049 seconds, a standard deviation of 0.039 seconds, and a maximum of 0.26 seconds.

$\proofAlg$ has no timeouts on the same 301 benchmarks, but overall it is slower than Z3 by a wide margin:  we presented its statistics for our selected benchmarks in Section~\ref{sec:proof-gen-eval}.
On the other hand, $\equivAlg$ performs comparably to Z3 on our selected benchmarks:  it has no timeouts, an average of 0.039 seconds, a median of 0.0074 seconds, a standard deviation of 0.11 seconds, and a maximum of 1.26 seconds.
The fact that Z3 hits the time limit on 8 benchmarks for which $\equivAlg$ and $\proofAlg$ do not hit the time limit does not necessarily imply that Z3 has worse coverage in general.
We also ran Z3 on the 21 regular expression pairs for which $\equivAlg$ hits the time limit (Section~\ref{sec:benchmarks}).
Z3 can prove all of them equivalent in less than two seconds each with proof generation enabled.
With proof generation disabled, each one takes less than a second.
Additionally, we ran Z3 on the 72 regular expression pairs for which $\proofAlg$ hits the time limit but $\equivAlg$ does not.
Z3 succeeds on each one in less than a second with proof generation enabled, and it succeeds on each one in less than half a second without proof generation enabled.

The fact that Z3 can generate proof certificates far more quickly than $\proofAlg$ can does not make it a viable substitute for $\proofAlg$.
As we explained in Section~\ref{sec:regex-equivalence}, the proof certificates that Z3 provides are not usable in ZK because they contain black-box term rewriting steps.

\subsection{Proof Validation Time}
\label{sec:eval-validation}

To answer $\researchQ2$, $\researchQ3$, and $\researchQ4$,
we use \tool to validate the proofs that we generate for our chosen benchmarks.
We run \tool with three different configurations for multiplexing:

\begin{enumerate}
\item\textbf{Default Settings.}
By default, \tool uses the multiplexing categories that we define in Section~\ref{sec:zk implementation}.
\item\textbf{No Multiplexing.}
The only checking instruction executed on a proof step is the one for that step's real underlying proof rule.
This causes \tool to leak information about the frequencies of different rules in the same way that ZKSMT does~\cite{luick2023zksmt}.
\item\textbf{Full Multiplexing.}
All rules except Assume and Contra (Appendix~\ref{sec:rules-extra}, Table~\ref{tab:assume-contra}) are grouped in a single category for multiplexing.
With full multiplexing, \tool leaks less information than it does under the default settings.
\end{enumerate}

\begin{figure}[!t]
 \centering
\ifsubmission
\else\includegraphics[width=0.95\textwidth]{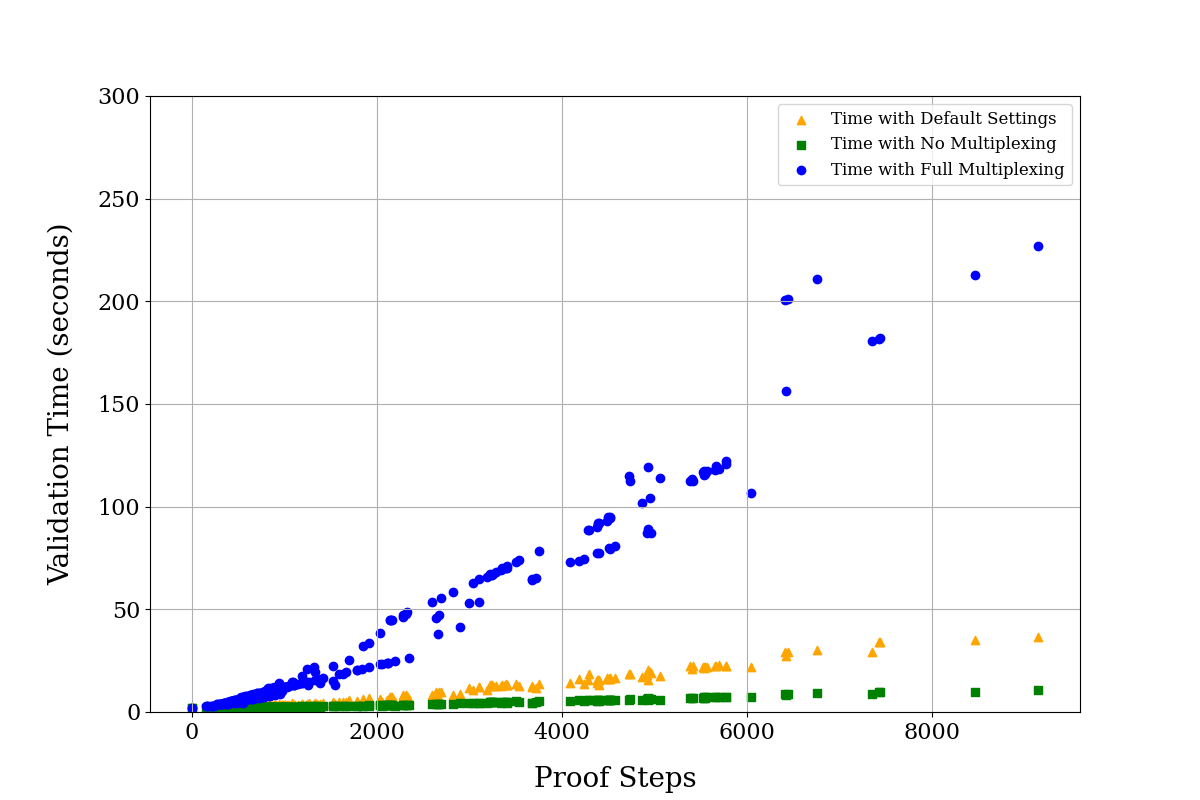}\fi
\caption{
Proof validation times for all configurations of \tool relative to proof step count.
}
\label{fig:validation}
\end{figure}

We tried other multiplexing configurations, such as using two multiplexing categories where one category contains all constant-time rules and the other contains all linear-time rules.
However, our preliminary results for them were inferior to the default settings.

\paragraph{Results}

We ran each of the three versions of \tool on all 301 of our selected benchmarks.
A scatter plot with our results for proof validation appears in Figure~\ref{fig:validation}.
We imposed a ten-minute timeout again for validation, but no benchmarks hit the time limit.
In general, the running time of the default version of \tool increases at a roughly linear rate as the number of steps in the proof increases.
The version with no multiplexing follows a similar pattern but is faster in most cases.
Full multiplexing is significantly slower than both of the other options.
The difference between it and the other two options widens as the number of proof steps increases.
The default version of \tool validates 84.39 percent of the benchmarks (254 out of 301) in 15 seconds or less.

With the default settings, the average validation time is 7.03 seconds,
the median is 3.10 seconds,
and the standard deviation is 7.27 seconds.
With no multiplexing, the average is 3.30 seconds,
the median is 2.27 seconds,
and the standard deviation is 1.87 seconds.
With full multiplexing, the average is 33.93 seconds,
the median is 10.71 seconds,
and the standard deviation is 44.58 seconds.
The slowest individual validation takes 36.64 seconds with the default settings, 10.75 seconds with no multiplexing, and 226.85 seconds with full multiplexing.
In the timing results for full multiplexing, there are two visibly distinct linear regressions.
Having a higher value of $\maxSyncLength$ causes some benchmarks to be slower relative to their step count.

The median ratio of \tool's running time with no multiplexing to its running time with the default settings is 0.74, so the added security that comes from multiplexing does not require a prohibitively large increase in running time.
The median ratio of \tool's running time with full multiplexing to its running time with the default settings is 3.42.

\section{Related Work}
\label{sec:related}

\paragraph{ZK Regular Expression Protocols}

Cryptographic protocols have been developed in the past for reasoning about regular expressions, but no existing protocol targets the problem that \tool does.
Instead of checking whether two regular expressions are equivalent, they check whether an individual string appears in the language of a regular expression or whether the string contains a substring that matches the regular expression.

Zombie~\cite{zhang2024zombie} can prove in zero knowledge that a packet satisfies network middlebox constraints involving Boolean combinations of regular expressions.
The text of the packet is private, and the constraints are public.
ZK-regex~\cite{luo2023privacy}
also provides ZK proofs that a private string matches a public regular expression.
Unlike Zombie, ZK-regex supports a two-party option for string matching where the string and regular expression are both private and belong to different parties.
The protocol zkreg~\cite{raymond2023efficient} provides succinct ZK proofs that a hidden string belongs to a public regular language.
Reef~\cite{angel2023reef} is another ZK protocol for matching private strings against public regular expressions that generates succinct proofs.
Unlike Zombie, ZK-regex, and zkreg, Reef supports features that are not allowed in pure regular expressions, such as lookahead assertions.

\tool doubles as a protocol for encoding string matching proofs:
to demonstrate that a private string $s$ matches a public regular expression $r$,
the prover can treat $s$ as a regular expression,
leak $r$ to the verifier,
and encode a proof that $s\regexUnion r=r$.
However, \tool would not be as efficient as Zombie, ZK-regex, zkreg, and Reef are because we designed it to handle a more general problem.

\paragraph{ZK Proof Validation}
ZKSMT~\cite{luick2023zksmt} is a ZK protocol for encoding SMT proofs.
It could be extended to encode regular expression equivalence proofs, but it is not as well-suited for the task as \tool is.
ZKSMT aims to provide a generalizable framework for proof validation in ZK.
To support arbitrary first-order theories, ZKSMT relies on list-based operations and type checking, which are costly operations for the protocol.
Neither one is necessary when we restrict ourselves to reasoning about only regular expressions.
Furthermore, ZKSMT provides weaker security guarantees than \tool does.
ZKSMT does not multiplex over rules:
it leaks the frequencies of different proof rules but not their order.
To our knowledge, there has been no prior work on reconstructing hidden proofs from the frequency distributions of their rules,
but ZKSMT does not provide a formal guarantee of its own security in that regard.

The protocol zkPi encodes proofs written in Lean, an interactive theorem prover~\cite{cryptoeprint:2024/267}.
In principle, zkPi could encode regular expression equivalence proofs.
However, Lean theorems need to be proven manually, and most programmers do not have the experience necessary to prove equivalences between regular expressions in Lean.
Also, zkPi supports a full calculus of dependent types that is unnecessary for regular expression equivalence proofs.

ZKUNSAT~\cite{luo2022proving} encodes proofs that pure Boolean formulas in conjunctive normal form are unsatisfiable.
It is inadequate for regular expression equivalence proofs as well.
Boolean satisfiability is in a lower complexity class than regular expression equivalence is, so the conversion from regular expressions to Boolean formulas would require a substantial increase in the input size.

\paragraph{General-Purpose ZK Protocols}

Protocols such as Cheesecloth~\cite{cuellar2023cheesecloth}, TinyRAM~\cite{ben2013snarks}, Pantry~\cite{braun2013verifying}, and Buffet~\cite{wahby2014efficient} can model the execution of arbitrary programs in zero knowledge.
However, general-purpose ZK protocols are highly inefficient in practice:  to validate a simple SMT proof, Cheesecloth needs to run for almost two hours~\cite{luick2023zksmt}.
Other general-purpose ZK protocols suffer from the same performance issues that Cheesecloth does, so they would be impractical for validation of regular expression equivalence proofs.

\paragraph{ZK Multiplexing}
More efficient implementations of ZK multiplexing exist than the version that \tool uses.
When \tool multiplexes over proof rules, it runs the rules' checking instructions sequentially.
Consequently, the asymptotic cost of multiplexing for \tool is linear in the number of proof rules within a multiplexing category.
When zkPi~\cite{cryptoeprint:2024/267} multiplexes over proof rules, it pays the cost of the most complex individual rule within a multiplexing category.
Other existing techniques for ZK multiplexing achieve a similar result~\cite{yang2023batchman}.
The Tight ZK CPU, a newer approach for ZK multiplexing, pays only the cost of the actual instruction used, not the most complex one~\cite{yang2024tight}.
\tool's ZK backend does not support any of these improved versions of ZK multiplexing currently, but \tool does use some optimizations to eliminate redundant operations without increasing its information leakage.

\paragraph{Coinductive Proofs}

Our work is not the first to apply coinduction to Kleene algebra.
A decision procedure for NetKAT program equivalence based on coinduction appears in~\cite{foster2015coalgebraic}.
The procedure uses state machine conversion, and it produces only a Boolean result rather than a proof tree.

An axiomatization of regular expression equivalence that resembles our calculus of rules appears in~\cite{corradini2002equational}.
Like our calculus, the axiomatization relies on derivatives and equations rather than state machine conversion.
However, there is no accompanying decision procedure for generating proofs with the provided axioms.

\paragraph{PSPACE-Complete Problems}
A quantified Boolean formula (QBF) is a Boolean formula that can contain universal and existential quantifiers applied to Boolean variables.
QBF satisfiability is a PSPACE-complete problem, just like regular expression equivalence is~\cite{giunchiglia2009reasoning}.
Moreover, all problems in PSPACE are solvable in ZK by reduction to QBF solving, with a translation based on Turing machines~\cite{shamir1992ip}.
QBF solving is a well-studied problem~\cite{beyersdorff2021quantified},
but we choose not to translate regular expressions into quantified Boolean formulas for the same reason that we choose not to translate them into state machines:
we would need to prove the correctness of the translation.

\paragraph{ReDoS Prevention}

The regular expression equivalence proofs that \tool encodes can help with the prevention of denial-of-service attacks, but, on its own, \tool itself does not prove that a regular expression is immune to denial-of-service attacks.
The task of checking whether an NFA is ambiguous is decidable in polynomial time~\cite{weber1991degree}.
There has been some prior research on the detection of DoS vulnerabilities in regular expressions~\cite{kirrage2013static,li2021redoshunter} and on the generation of regular expressions that are immune to DoS attacks~\cite{li2020flashregex}, but the task of providing proof certificates to confirm that a regular expression or NFA is not vulnerable to denial-of-service attacks has not received any significant attention.

To eliminate a ReDoS vulnerability, one can either repair the vulnerable regular expression directly or switch to a different string matching algorithm more suitable for the domain at hand.
\tool validates regular expression modifications, not algorithmic changes.
Cloudflare addressed their vulnerability by switching to DFA-based matching~\cite{cloudflare2020}.
Although Cloudflare made an algorithmic change, the availability of DFA-based matching algorithms does not trivialize the problem of ReDoS prevention:  in general, a DFA can be exponentially larger than its corresponding NFA.

\section{Conclusion}
\label{sec:conclusion}

We have introduced \tool, the first ZK protocol designed to support regular expression equivalence proofs and the first ZK protocol to target a PSPACE-complete problem.
Multiple potential directions exist for
future work based on \tool and its custom calculus of proof rules.
The integration of $\proofAlg$ or a similar algorithm into mainstream SMT solvers would facilitate the use of regular expression equivalence proofs in practice.

\tool can be modified to support proofs about Kleene Algebra with Tests (KAT).
KAT extends the structure of regular expressions with variables and logical connectives~\cite{kozen1997kleene}.
KAT itself supports further extensions with theories such as bit vectors~\cite{greenberg2022kleene}, the behavior of packets in a network~\cite{anderson2014netkat}, and Hoare logic~\cite{kozen2002hoare,antonopoulos2023algebra}.
For this paper, we choose to focus on ordinary regular expressions because they are the most common application of Kleene algebra in practice.

In \tool's model, both private regular expressions belong to the same party.
An additional direction for future work would be the development of a privacy-preserving regular expression equivalence protocol for the scenario where the regular expressions being compared belong to different parties.
Unlike \tool, the multi-party protocol would need to prove the regular expressions' equivalence in a cryptographic setting rather than relying on a proof constructed offline.

\paragraph{Limitations}
We do not consider intersections, complements, backreferences, or lookahead assertions for \tool since they are not allowed in pure regular expressions.
It is possible to extend \tool's calculus of rules to support them, but we would lose decidability for equivalence if we incorporated backreferences or lookahead assertions~\cite{freydenberger2013extended}.

\bibliographystyle{plain}
\bibliography{acmart}

\appendix

\section{Simple Proof Rules}
\label{sec:rules-extra}

\begin{table}[!t]
    \centering
    \begin{tabular}{|c|c|}
    \hline 
    {\sf RuleID} & {\sf Conclusion}\\    
    \hline
    UnionAssoc
    &
    $p\regexUnion(q\regexUnion r)=(p\regexUnion q)\regexUnion r$
    \\
    \hline

    UnionComm
    &
    $p\regexUnion q=q\regexUnion p$
    \\
    \hline

    UnionEmpty
    &
    $p\regexUnion\regexEmpty=p$
    \\
    \hline  

    UnionSelf
    &
    $p\regexUnion p=p$
    \\
    \hline
    
    ConcatAssoc
    &
    $p(qr)=(pq)r$
    \\
    \hline

    ConcatBlankL
    &
    $\regexBlank p=p$
    \\
    \hline
    
    ConcatBlankR
    &
    $p\regexBlank=p$
    \\
    \hline
    
    ConcatEmptyL
    &
    $\regexEmpty p=\regexEmpty$
    \\
    \hline

    ConcatEmptyR
    &
    $p\regexEmpty=\regexEmpty$
    \\
    \hline
    
    \end{tabular}
    \caption{Normalization Rules}
    \label{tab:normalization}
\end{table}

\paragraph{Normalization Rules}

The normalization rules appear in Table~\ref{tab:normalization}.
None of the normalization rules have premises.
Note that we do not include all of the standard axioms of Kleene algebra as proof rules~\cite{kozen1997kleene}.
There are no rules for manipulating stars or for distributing concatenations over unions.
We can still uphold completeness with the axioms that we have, so we omit the others to avoid unnecessary complexity.

In the ZK proofs that $\proofAlg$ produces as output, we need to confirm that our transformations for normalization are valid.
However, we never need to prove that the end result of normalization is a regular expression in normal form.
Whether a regular expression is in normal form does not have any bearing on which proof rules can be applied to it:  its only importance is that it forces $\proofAlg$ to terminate.

\begin{table}[!t]
    \centering
    \begin{tabular}{|c|c|c|}
    \hline 
    {\sf RuleID} & {\premise} & {\sf Conclusion}\\    
    \hline

    Refl
    &
    &
    $x=x$
    \\
    \hline
    
    Symm
    &
    $x=y$
    &
    $y=x$
    \\
    \hline
    
    Trans
    &
    $x=y$, $y=z$
    &
    $x=z$
    \\
    \hline

    PredCongL
    &
    $P(x_1, y)$, $x_1=x_2$
    &
    $P(x_2, y)$
    \\
    \hline

    PredCongR
    &
    $P(x, y_1)$, $y_1=y_2$
    &
    $P(x, y_2)$
    \\
    \hline

    FunCong1
    &
    $x_1=x_2$
    &
    $f(x_1)=f(x_2)$
    \\
    \hline

    FunCong2
    &
    $x_1=x_2$, $y_1=y_2$
    &
    $f(x_1, y_1)=f(x_2, y_2)$
    \\
    \hline
    
    \end{tabular}
    \caption{Equality Rules}
    \label{tab:equality}
\end{table}

\paragraph{Equality Rules}

The equality rules appear in Table~\ref{tab:equality}.
All of our equality rules are standard axioms of first-order logic.
The rules Refl, Symm, and Trans simply allow us to use equality as an equivalence relation.
Our congruence rules only need to support binary predicates because
all predicates in our formalism take two terms as arguments except $\bot$, which takes none.
Sync also takes a string as an argument, but we never need to perform substitutions for strings.

\begin{table}[!t]
    \centering
    \begin{tabular}{|c|c|c|}
    \hline 
    {\sf RuleID} & {\premise} & {\sf Conclusion}\\
    \hline

    EpsilonEmpty
    &
    &
    $\hasblank\regexEmpty=\regexEmpty$
    \\
    \hline
    
    EpsilonBlank
    &
    &
    $\hasblank\regexBlank=\regexBlank$
    \\
    \hline
    
    EpsilonChar
    &
    &
    $\hasblank c=\regexEmpty$
    \\
    \hline

    EpsilonUnionPos1
    &
    $\hasblank{p}=\regexBlank$
    &
    $\hasblank{p\regexUnion q}=\regexBlank$
    \\
    \hline

    EpsilonUnionPos2
    &
    $\hasblank{q}=\regexBlank$
    &
    $\hasblank{p\regexUnion q}=\regexBlank$
    \\
    \hline

    EpsilonUnionNeg
    &
    $\hasblank{p}=\regexEmpty$, $\hasblank{q}=\regexEmpty$
    &
    $\hasblank{p\regexUnion q}=\regexEmpty$
    \\
    \hline

    EpsilonConcatPos
    &
    $\hasblank p=\regexBlank$, $\hasblank q=\regexBlank$
    &
    $\hasblank{pq}=\regexBlank$
    \\
    \hline

    EpsilonConcatNeg1
    &
    $\hasblank p=\regexEmpty$
    &
    $\hasblank{pq}=\regexEmpty$
    \\
    \hline

    EpsilonConcatNeg2
    &
    $\hasblank q=\regexEmpty$
    &
    $\hasblank{pq}=\regexEmpty$
    \\
    \hline

    EpsilonStar
    &
    &
    $\hasblank{p^*}=\regexBlank$
    \\
    \hline
    
    \end{tabular}
    \caption{Epsilon Rules}
    \label{tab:epsilon}
\end{table}

\paragraph{Epsilon Rules}

Our epsilon rules appear in Table~\ref{tab:epsilon}.
Instead of having a single rule for the epsilon function, we have a separate rule for each regular expression node.
The definition for each case is standard~\cite{brzozowski1964derivatives}.

\begin{table}[!t]
    \centering
    \begin{tabular}{|c|c|}
    \hline 
    {\sf RuleID} & {\sf Conclusion}\\    
    \hline

    DeriveEmpty
    &
    $\derivative c\regexEmpty=\regexEmpty$
    \\
    \hline
    
    DeriveBlank
    &
    $\derivative c\regexBlank=\regexEmpty$
    \\
    \hline
    
    DeriveCharSame
    &
    $\derivative cc=\regexBlank$
    \\
    \hline

    DeriveCharDifferent
    &
    $\derivative{c_1}{c_2}=\regexEmpty$ if $c_1$ is not $c_2$
    \\
    \hline

    DeriveUnion
    &
    $\derivative c{p\regexUnion q}=\derivative cp\regexUnion\derivative cq$
    \\
    \hline

    DeriveConcat
    &
    $\derivative c{pq}=\derivative cpq\regexUnion\hasblank p\derivative cq$
    \\
    \hline

    DeriveStar
    &
    $\derivative c{p^*}=\derivative cpp^*$
    \\
    \hline
    
    \end{tabular}
    \caption{Derivative Rules}
    \label{tab:derivative}
\end{table}

\paragraph{Derivative Rules}

The derivative rules appear in Table~\ref{tab:derivative}.
They follow the standard definition of the derivative function~\cite{brzozowski1964derivatives}.
DeriveCharDifferent is the only rule in our calculus with a \textit{side condition}.
The side condition is not a premise, nor is it part of the conclusion.
It does not appear in any table at all.
The checking instruction for DeriveCharDifferent merely checks at runtime that the AST nodes for $c_1$ and $c_2$ have different characters.
The conclusion of DeriveCharDifferent that appears in $\formulaTable$ is simply $\derivative{c_1}{c_2}=\regexEmpty$.

\begin{table}[!t]
    \centering
    \begin{tabular}{|c|c|c|}
    \hline 
    {\sf RuleID} & {\premise} & {\sf Conclusion}\\    
    \hline

    Assume
    &
    &
    $p\neq q$
    \\
    \hline
    
    Contra
    &
    $p\neq q$, $p=q$
    &
    $\bot$
    \\
    \hline
    
    \end{tabular}
    \caption{Proof Completion Rules}
    \label{tab:assume-contra}
\end{table}

\paragraph{Proof Completion Rules}

We need two additional rules to bookend our proofs:  Assume and Contra.
They appear in Table~\ref{tab:assume-contra}.
A proof contains one use of Assume as its starting point.
Assume can introduce only formulas of the form $p\neq q$, so it cannot trivialize a proof.
Contra concludes a proof:  it derives $\bot$ from $p\neq q$ and $p=q$.
Assume is the only rule that can introduce inequality formulas, so the first premise of Contra must come from Assume.

For multiplexing, Assume and Contra each have a category to themselves.
Proofs always contain exactly one occurrence of each, so this is not a source of information leakage.

\section{Rule Redundancy}

Match is redundant with Coinduction:  any equivalence that can be proven with Match can also be proven less directly with Coinduction and our auxiliary coinduction rules (Table~\ref{tab:unified}).
We choose to include Match in our calculus of rules anyway because applications of Match are more efficient to validate in ZK than applications of Coinduction are.
When we split Coinduction into $\alphabetsize$ steps, some of the steps take linear time in $\maxSyncLength$ to check.
When we split Match into $\alphabetsize$ steps, all of the steps take constant time to check.

\section{Complex ZK Checking Instructions}
\label{sec:instructions_in_zk}

Most of \tool's checking instructions are straightforward to execute in ZK:
they simply retrieve a fixed number of AST nodes from read-only memory and perform a fixed number of comparisons of committed values.
The checking instructions whose ZK implementations are non-trivial are the ones that run in linear time relative to $\maxSyncLength$.
For our linear-time rules, we need to reason about entire linked lists and, consequently, about multisets as well.

\paragraph{Derivative Chain and String Scanning}
For some of our proof rules,
namely SyncCycle, EqualSync, and Coinduction,
we need to confirm that a chain of nested derivatives matches a Sync string.
Our method for comparing derivative chains and Sync strings for these rules depends on our representation of derivative chains in $\termTable$.
Multi-character derivatives do not exist in $\termTable$:
for a regular expression $p$ and a string $s=s_1s_2\dots s_m$, $\derivative sp$ is stored as $\derivative{s_m}{\dots\derivative{s_2}{\derivative{s_1}{p}}}$.
The derivatives for characters at the end of the string are on the outside of the chain of nested derivatives, and the derivatives for characters at the start of the string are on the inside.
This is the opposite of the ordering used in $\syncTable$, where the AST node for $s_1$ has a pointer to $s_2$, which has a pointer to $s_3$, and so on until we reach $s_m$, which has a pointer to the null terminator.
Effectively, to check that we have the same string $s$ in $\syncTable$ as we do in the derivative chain, we need to confirm that one singly-linked list is the reversal of another.

Algorithm~\ref{alg:linked-list-reversal} is the procedure that we use to check that a derivative chain is the reversal of a Sync string.
For the rules that require it, the derivative chain and the Sync string must be of the same length and not more than $\maxSyncLength$ entries long.
In their checking instructions, we keep track of two multisets $A$ and $B$, the former for storing the contents of the derivative chain and the latter for storing the contents of the Sync string.
We cannot simply add the characters in each linked list to their respective multisets because we need to enforce that one linked list is the reversal of the other.
Instead, we encode each character as an integer, and we combine the character's integer with a numerical index that represents its AST node's position within a linked list.
We index the entries in the derivative chain in ascending order, and we index the entries in the Sync string in descending order.
At the end, we confirm that the two lists are equivalent when viewed as multisets by comparing $A$ and $B$.
The two multisets will be equivalent if and only if the derivative chain and the Sync string contain the same characters in opposite orders.
We know what value to use at the start for the descending Sync string indices because every entry in $\syncTable$ stores its height as a committed integer (line~\ref{line:get-height}).
The height of a string node is simply the node's distance from the null string terminator in terms of AST child pointers.

If the two linked lists are fewer than $\maxSyncLength$ entries long, we still run $\maxSyncLength$ loop iterations.
We pad $A$ and $B$ with zeroes rather than meaningful entries once we reach the end of the linked lists.
The variable $z$ on line~\ref{alg:line-fake-iteration} captures this.
Once the loop iterations exceed the true height of $s$, $z$ switches from 1 to 0, and the loop performs null operations for its remaining iterations.

On lines \ref{line:join-first} and \ref{line:join-second},
the function $\texttt{join}$ concatenates two integers together into a single committed integer.
Assuming that we know an upper bound on the values of both integers, we can combine them without loss of information by bit-shifting one integer and adding it to the other.

\SetKwComment{Comment}{/* }{ */}
\SetKwInput{KwProof}{Proof}
\SetKwInput{KwOutput}{Output}
\SetKwFunction{KwSubchecker}{{CheckingInstrs}}
\SetKwFunction{KwPermchecker}{PermuteCheck}
\begin{algorithm}[!t]
\caption{$\texttt{checkReverse}(t,s)$}\label{alg:linked-list-reversal}

$A,B\leftarrow[],[]$\;

$h\leftarrow\texttt{height}(s)$\;\label{line:get-height}

$\hat s,\hat t\leftarrow s, t$\;

$k\leftarrow1$\;

\While{$k\leq\maxSyncLength$}{
$z\leftarrow \texttt{int}(k\leq h)$\;\label{alg:line-fake-iteration}
$\derivative{c_a}{t'}\leftarrow \hat t$\;
$c_bs'\leftarrow \hat s$\;
$x_a\leftarrow z*\texttt{join}(k, c_a)$\;\label{line:join-first}
$x_b\leftarrow z*\texttt{join}(1+h-k, c_b)$\;\label{line:join-second}
$A,B\leftarrow A::x_a, B::x_b$\;
$\hat t\leftarrow z \texttt{ ? } \termTable[t'] \texttt{ : } \hat t$\;
$\hat s\leftarrow z \texttt{ ? } \syncTable[s'] \texttt{ : } \hat s$\;
$k\leftarrow k+1$\;
}

${\bf assert}(\hat s=\regexBlank)$\;

${\bf assert}(\mset{A}=\mset{B})$\;
\end{algorithm}

\paragraph{Sync String Scanning}
For the premises of Coinduction for individual characters, we need to check that the string $sc$ is identical to the string $s$ apart from the extra character $c$ that it has at the end.
Because we represent multi-character strings by adding extra characters to the front of shorter strings,
the two strings do not have any AST nodes in common except the null terminator at the end.
To compare $sc$ and $s$, we iterate over the two string simultaneously.
We start from the starting pointers for both strings.
In each iteration, we check that the characters at the current nodes are the same, and then we move to considering the immediate children of the two nodes we just compared.
At the end, we confirm that $s$ has reached the null terminator.
We also confirm that $sc$ has one additional character at the end and a null terminator after that.

When we compare $sc$ and $s$ in ZK, we pad the loop with extra iterations.  Once $s$ reaches the null terminator, we stop moving forward for either $s$ or $sc$ and simply compare the same nodes repeatedly until we reach $\maxSyncLength$ iterations.

\section{Completeness}
\label{sec:complete}

Let $p$ and $q$ be two equivalent regular expressions.
We want to show that there exists a proof that $p=q$ in our calculus of rules.
We will give an algorithm for finding the equivalence proof.
Before we begin the completeness proof, we will state a number of preliminary findings whose proofs we omit.

Our preliminary findings depend on the definition of similarity from Section~\ref{sec:proof-gen}:  two regular expressions are similar if one can be converted into the other using only our normalization rules and equality rules.
We write $p\similar q$ to denote that $p$ and $q$ are similar.
We also rely on the functions $\reduceFunction$ and $\normalizeFunction$ from the same section.

\begin{theorem}[Term Conversion]
\label{theorem:term-conversion-no-proof}
Any term $r$ can be converted into an equivalent regular expression using our normalization, equality, epsilon, and derivative rules.
In particular, if $r$ is $\hasblank{r'}$, then $r$ can be converted into either $\regexEmpty$ or $\regexBlank$, but not both.
\end{theorem}

\paragraph{Reduction Function}
Building on Theorem~\ref{theorem:term-conversion-no-proof},
let $\reduceFunction$ be a function that takes a term $r$ as input and produces an equivalent regular expression by unfolding any $\hasblankFunction$ and $\derivativeFunction$ nodes.
If $r$ is already a regular expression, $\reduce r$ returns $r$ itself.

\paragraph{Normal Form}
Let $\totalOrder$ be an arbitrary total order over terms.
We say that a term is in \textit{normal form} if it satisfies the following conditions according to its \nodeid:

\begin{enumerate}
\item $\regexEmpty$, $\regexBlank$, and individual characters are all in normal form.
\item $p\regexUnion q$ is in normal form if it satisfies one of two combinations of conditions:
\begin{enumerate}
\item $q$ is a union $q_1\regexUnion q_2$, $p$ and $q$ are in normal form, $p\neq q_1$, $p$ is not a union, neither $p$ nor $q_1$ is $\regexEmpty$, and $p<q_1$ according to $\totalOrder$.
\item $q$ is not a union, $p$ and $q$ are in normal form, $p\neq q$, $p$ is not a union, neither $p$ nor $q$ is $\regexEmpty$, and $p<q$ according to $\totalOrder$.
\end{enumerate}
\item $pq$ is in normal form if $p$ and $q$ are in normal form, $p$ is not a concatenation, and neither $p$ nor $q$ is $\regexEmpty$ or $\regexBlank$.
\item $p^*$ is in normal form if $p$ is.
\end{enumerate}

Note that terms in normal form must be regular expressions:  our requirements forbid derivative and epsilon nodes from appearing.

\begin{theorem}[Normal Form Conversion]
There exists a function $\normalizeFunction$ such that,
for any regular expression $r$, the output of $\normalize{r}$ is a regular expression in normal form that is similar to $r$.
\end{theorem}

\begin{theorem}[Uniqueness of Normal Form]
\label{theorem:normal-unique-no-proof}
A regular expression in normal form is not similar to any normal-form regular expression other than itself.
\end{theorem}

\subsection{Main Completeness Proof}

Let $p$ and $q$ be two equivalent regular expressions.
We want to construct a proof of $p=q$ using our calculus of rules.

To start, let $P$ be the set of all regular expressions of the form $\normalize{\reduce{\derivative sp}}$ for strings $s$.
Since $s$ can be the empty string, $P$ contains the normalized version of $p$ itself.
Following the same pattern, let $Q$ be the set of all regular expressions of the form $\normalize{\reduce{\derivative sq}}$.
We know from \cite{brzozowski1964derivatives} that the set of all unfolded derivatives for $p$ and $q$ must fall into a finite number of equivalence classes for similarity.
Because the regular expressions in $P$ and $Q$ are all normalized, combining that result with Theorem~\ref{theorem:normal-unique-no-proof} gives us that $P$ and $Q$ must be finite sets.

Let $k=|P|\cdot|Q|$, and consider the set $K$ of all strings in $\alphabet$ of length $k$.
A string $s\in K$ has $k+1$ distinct prefixes, including both the empty string and $s$ itself.
If we view similarity as a relation over pairs of regular expressions,
the set $P\times Q$ has at most $k$ distinct similarity classes, so there must be at least two prefixes of $s$ whose derivatives for $p$ and $q$ are similar.
Let $\head s$ and $\foot s$ be the shortest two distinct prefixes of $s$ such that $\derivative{\head s}p\similar\derivative{\foot s}p$ and $\derivative{\head s}q\similar\derivative{\foot s}q$, where $\head s$ is shorter than $\foot s$.
Also, let $\body s$ be the non-empty string such that $\foot s=\head s\body s$.
We can use SyncCycle to derive $\Sync{\body s}{\derivative{\head s}p}{\derivative{\head s}q}$.
Next, we can apply SyncFold repeatedly to convert that formula into $\Sync{\foot s}pq$.

For the next step, let $F_0$ be the set of all formulas of the form $\Sync{\foot s}pq$ that can be constructed from strings in $K$ using this procedure.
Note that two different strings in $K$ may map to the same element of $F_0$.
$F_0$ is the initial \textit{frontier} for our proof generation algorithm.
A frontier $F$ must uphold a number of invariants:
\begin{enumerate}
\item\label{inv:finite} $F$ is a finite set.
\item\label{inv:sync} Every formula in $F$ is of the form $\Sync upq$ for some string $u$.
\item\label{inv:conflict} If $s_1$ and $s_2$ are strings such that the formulas $\Sync{s_1}pq$ and $\Sync{s_2}pq$ are both in $F$, then $s_1$ is not a strict prefix of $s_2$.
\item\label{inv:coverage} For every string $s$ of length $k$, $F$ contains $\Sync tpq$ for some string $t$ that is a non-strict prefix of $s$.
\item\label{inv:proof} Every formula in $F$ has a proof in our calculus of rules.
\end{enumerate}

All of these hold trivially for our initial frontier except invariant~\ref{inv:conflict}.
We can prove invariant~\ref{inv:conflict} for $F_0$ by contradiction.
Assume that we have both $\Sync{\foot{s_1}}pq$ and $\Sync{\foot{s_2}}pq$ in $F_0$, and, without loss of generality, assume that $\foot{s_1}$ is a strict prefix of $\foot{s_2}$.
According to our definition from before, $\foot{s_1}$ is the shortest prefix of $s_1$ that hits a cycle for similarity.
If $\foot{s_1}$ is a prefix of $\foot{s_2}$, then $\foot{s_1}$ is a prefix of $s_2$ as well.
This would make $\foot{s_1}$ a prefix of $s_2$ that hits a cycle for similarity.
$\foot{s_1}$ is strictly shorter than $\foot{s_2}$, so this contradicts the definition of $\foot{s_2}$ as the shortest prefix of $s_2$ that hits a cycle for similarity.
Therefore, this situation is impossible, and $F_0$ must uphold invariant~\ref{inv:conflict}.

Now we need to define an algorithm for constructing a proof.
We will define a function $\update F$ that takes a frontier $F$ and produces a new frontier.
To start, let $m$ be the maximum depth of any entry of $F$.
We define the depth of a formula $\Sync upq$ as the length of the string $u$ that it contains.
If $m$ is 0, return $F$.
If $m$ is positive, select an element $\Sync wpq$ from $F$ such that the string $w$ is of length $m$.
We know that $w$ is non-empty, so let $c$ be the final character of $w$, and let $w'$ be all of the characters before it.
Let $B$ be the set of all formulas of the form $\Sync{wb}pq$ for characters $b\in\alphabet$.
Let $F'=(F\cup\{\Sync wpq\})\setminus B$.
Return $F'$.

If $F$ is a valid frontier, then so is $F'$.
Assume that $F$ upholds all of the required invariants.
Invariants~\ref{inv:finite} and~\ref{inv:sync} are simple to prove.
$F'$ must be finite because only one formula appears in $F'$ that does not also appear in $F$.
$F$ is finite, so $F'$ must be finite as well.
Also, every formula in $F'$ is of the correct form because the only formula in $F'$ that does not also appear in $F$ is $\Sync wpq$.

$F'$ upholds invariant~\ref{inv:conflict} because $F$ does.
The only new entry in $F'$ is $\Sync wpq$.
Suppose that we have strings $s_1$ and $s_2$, where $s_1$ is a strict prefix of $s_2$, such that $\Sync{s_1}pq$ and $\Sync{s_2}pq$ are in $F'$.
If neither $s_1$ nor $s_2$ is $w$, then the pair is also present in $F$, which is impossible.

If $s_1$ is $w$, then $s_2$ must be more than one character longer than $w$ because $\Sync{wb}pq$ does not appear in $F'$ for any character $b$.
Besides those formulas and $\Sync wpq$, every formula in $F'$ also appears in $F$, so $\Sync{s_2}pq$ must appear in $F$ as well.
Let $c$ be a character and $t$ be a non-empty string such that $s_2=wct$.
$F$ contains the formula $\Sync{wc}pq$, so we have a prefix conflict between the formulas $\Sync{wc}pq$ and $\Sync{wct}pq$ in $F$, which is impossible.

If $s_2$ is $w$, then $\Sync{s_1}pq$ must appear in $F$.
Since $s_1$ is a strict prefix of $w$, it must be a strict prefix of $wc$ as well for any character $c$.
We know that $\Sync{wc}pq$ appears in $F$, so we have a prefix conflict between $\Sync{s_1}pq$ and $\Sync{wc}pq$ in $F$, which is impossible.
This eliminates all of the possible ways that $F'$ could violate invariant~\ref{inv:conflict}.

$F'$ upholds invariant~\ref{inv:coverage} because, for every $b\in\alphabet$, the formula $\Sync wpq$ functions as a replacement for $\Sync{wb}pq$.
The string $w$ is a prefix of $wb$, so any string $s$ of length $k$ that was covered by $\Sync{wb}pq$ in $F$ is covered by $\Sync wpq$ in $F'$.

For invariant~\ref{inv:proof}, we need to confirm that there exists a valid proof of $\Sync wpq$, given that, for every $b\in\alphabet$, there exists a valid proof of $\Sync{wb}pq$.
To get a proof of $\Sync wpq$, we can apply our Coinduction rule, taking the proofs of $\Sync{wb}pq$ for every $b\in\alphabet$ as premises.
As an additional premise, we also need a proof that $\hasblank{\derivative wp}=\hasblank{\derivative wq}$.
The proof must exist because of Theorem~\ref{theorem:term-conversion-no-proof}.

Now that we have defined $\update{F}$, we need to define our algorithm for using it.
We start from the initial frontier $F_0$.
Apply the $\updateFunction$ function to $F_0$ repeatedly until the maximum depth of a formula in the frontier is 0.
We know that this loop will terminate eventually.
Let $F'=\update F$ for some frontier $F$.
We know from the definition of $\updateFunction$ that $F'$ is created from $F$ by adding $\Sync wpq$ and removing all formulas of the form $\Sync{wb}pq$ for some string $w$ and all characters $b\in\alphabet$.
The string $w$ is shorter than $wb$ for any $b$, so adding $w$ to the set does not increase the maximum depth.
Also, note that $F'$ removes $\alphabetsize=|\alphabet|$ formulas of some depth $m$ and adds a single formula of depth $m-1$.
The frontier must be a finite set, so the formulas of depth $m$ must be depleted eventually.
Once the formulas of depth $m$ are eliminated, the loop moves to eliminating the formulas of depth $m-1$, $m-2$, and so on until the formulas in the frontier are all of depth 0.

\ifsubmission
\else
There is only one Sync formula of depth 0 that a frontier can contain, namely $\Sync{\regexBlank}pq$.
\fi
Once we have $\Sync{\regexBlank}pq$ in the frontier, we can apply SyncEmpty to reach the conclusion that $p=q$.
When we reach the end, we know that a valid proof exists for $\Sync{\regexBlank}pq$, so adding an application of SyncEmpty gives us a valid proof of $p=q$.

Since we can always prove $p=q$ for a pair of equivalent regular expressions using our calculus of rules, we can always make a proof that \tool accepts.
If we have a proof of $p=q$, we can apply Assume to derive $p\neq q$ and then apply Contra to derive $\bot$.

\section{Plaintext Validation}

For all three validation configurations that we examine,
the time cost of running \tool comes almost entirely from our ZK operations rather than the structure of the algorithm itself.
If we execute \tool on our 301 selected benchmarks with no ZK operations, it runs much more quickly.
With all ZK operations disabled, the benchmarks have an average running time of 0.0049 seconds and a median running time of 0.0033 seconds.
The standard deviation is 0.0030 seconds.
The slowest individual benchmark takes only 0.017 seconds to validate.
For comparison, the fastest individual ZK benchmark takes 1.74 seconds.

\end{document}